\documentclass[proof]{pasj00}

\SetRunningHead{Y. Hasegawa and T. Tsuribe}{Possibility of Kelvin-Helmholtz Instability}
\Received{2012/09/18}
\Accepted{2012/12/11}
\begin{document}
\title{The Possibility of the Kelvin-Helmholtz Instability during Sedimentation of Dust Grains in the Protoplanetary Disk}
\author{Yukihiko \textsc{Hasegawa}, and Toru \textsc{Tsuribe}}
\affil{Department of Earth and Space Science, Graduate School of Science, Osaka University, Toyonaka, Osaka 560-0043}
\email{hasegawa@vega.ess.sci.osaka-u.ac.jp, tsuribe@vega.ess.sci.osaka-u.ac.jp}
\maketitle

\begin{abstract}
In this paper, we reexamine the possibility of shear-driven turbulence during sedimentation of dust grains in the protoplanetary disk. The shear-driven turbulence is expected to occur before the onset of the gravitational instability for MMSN model. While according to previous studies without taking account of growth of dust grains, with the larger abundance of dust grains, the gravitational instability is indicated to occur before shear-driven turbulence. In this paper, the case with dust growth is considered, and it is found that the Kelvin-Helmholtz instability tends to occur before the gravitational instability even in the case with large abundance of dust grains. This is different from previous results without the dust growth.
\end{abstract}

\section{Introduction}

For the formation of planetesimals, growth and motion of dust grains in the protoplanetary disks are essential. In a typical scenario, dust grains settle toward the midplane, and a thin dust layer is supposed to form (\cite{key-N81}; \cite{key-N86}). The gravitational instability (GI) of a disk (\cite{key-S69}; \cite{key-G73}; \cite{key-S83}) is expected to occur when density in the disk exceeds the critical density that is given by
\begin{equation}
\rho _\mathrm{c} = \frac{0.606 \MO}{r^3} \mathrm{,} \label{eq:01}
\end{equation}
where $r$ is the heliocentric distance (\cite{key-S83}, 1998). In order to resolve the problem of radial drift of meter-sized dust (\cite{key-A76}), GI in the thin dust layer is essential. However, as dust grains settle toward the midplane, vertical dust density gradient increases. In such a case, the vertical shear of the rotational velocity in the dust layer becomes strong. The strong shear is expected to induce the Kelvin-Helmholtz instability (KHI) (\cite{key-C61}). As a result, KHI is expected to induce shear-driven turbulence. It is considered that turbulence due to KHI prevents dust grains from settling further toward the midplane (\cite{key-S98}; \cite{key-S01}; \cite{key-M06}). \citet{key-S98} shows that GI does not occur in MMSN (minimum mass solar nebula) model (\cite{key-H81}; \cite{key-H85}). On the other hand, \citet{key-S98} and \citet{key-S01} show that turbulence induced by KHI becomes weaker if dust abundance in the protoplanetary disk is much larger than MMSN model. If turbulence is weak, GI will occur and planetesimals may form. To discuss the possibility of the shear-driven turbulence, Richardson number is known as an indicator of KHI (\cite{key-C61}). If the Richardson number is smaller than a critical value, KHI is expected to occur. \citet{key-C61} showed that the critical value is 0.25, but \citet{key-G05} and \citet{key-J06} showed that the inclusion of the Coriolis force yields a much higher critical Richardson numbers. Other than KHI, the dynamics of the midplane has been shown to be dominated by the streaming instability (\cite{key-Y05}; \cite{key-J07}).

Investigations of the turbulence are essential for understanding the processes of the planetesimal formation, and there are many previous studies (\cite{key-W80}; \cite{key-C93}; \cite{key-D99}; \cite{key-J07}; \cite{key-B10}). These previous studies confirm the result that shows that GI is expected to occur before KHI if dust abundance is large. However, above previous studies did not take account of dust growth. Not only gas turbulence but also dust growth are essential to understand the planetesimal formation. Dust grains grow due to dust-dust collisions while they settle toward the midplane (\cite{key-N81}; \cite{key-N86}). \citet{key-N86} shows that the law of gas drag force on dust grains is changed in the process of sedimentation of dust grains because of dust growth. They also show that the change of the law of gas drag influences sedimentation of dust grains. Dust growth is strongly related to gas drag. However, it is difficult to simulate both dust growth and gas turbulence numerically because of the computer performance.

In this paper, we numerically calculate the sedimentation and the growth of dust grains, and the possibility of KHI is discussed using the distribution of dust density that is consistent with their sedimentation in the disk. In \S 2, the models of gas and dust grains of the protoplanetary disk in this paper are summarized. In \S 3, we show numerical results for sedimentation of dust grains without growth. In \S 4, results are shown for the case with dust growth. In \S 5, we discuss the effects neglected in this paper. In \S 6, we summarize our results.

\section{Models of the protoplanetary disk}

\subsection{Disk model}

For simplicity, we restrict ourselves to the case with $r = 1$ AU. The protoplanetary disk is composed of gas and dust. In this paper, the gas surface density $\Sigma _\mathrm{g}$ and the dust surface density $\Sigma _\mathrm{d}$ are assumed to be
\begin{equation}
\Sigma _\mathrm{g} = 1.7 \times 10^3 f_\mathrm{g} \left( \frac{r}{1 [\mathrm{AU}]} \right)^{-\frac{3}{2}} [\mathrm{g} \, \mathrm{cm}^{-2}] \mathrm{,} \label{eq:02}
\end{equation}
\begin{equation}
\Sigma _\mathrm{d} = 7.1 f_\mathrm{d} \, \xi _\mathrm{ice} \left( \frac{r}{1 [\mathrm{AU}]} \right)^{-\frac{3}{2}} [\mathrm{g} \, \mathrm{cm}^{-2}] \mathrm{,} \label{eq:03}
\end{equation}
where $f_\mathrm{g}$, $f_\mathrm{d}$ and $\xi _\mathrm{ice}$ are parameters of abundance for gas, dust and condensed water ice, respectively (\cite{key-H81}; \cite{key-H85}). At $r = 1$ AU, $\xi _\mathrm{ice} = 1$. The case with $f_\mathrm{g} = 1$ and $f_\mathrm{d} = 1$ corresponds to MMSN model. In this paper, we consider the case with $f_\mathrm{g} = 1$ and various $f_\mathrm{d}$. In MMSN model, temperature $T$ is given by
\begin{equation}
T = 280 \left( \frac{r}{1 [\mathrm{AU}]} \right)^{-\frac{1}{2}} [\mathrm{K}] \mathrm{.} \label{eq:04}
\end{equation}

\subsection{Gas properties}

In this paper, we assume that the gas component is in hydrostatic equilibrium in the vertical direction without self-gravity. In this case, the gas density is given by
\begin{equation}
\rho _\mathrm{g}(z) = \frac{\Sigma _\mathrm{g}}{\sqrt{\pi } H_\mathrm{g}} \exp \left[ - \left( \frac{z}{H_\mathrm{g}} \right)^2 \right] \mathrm{,} \label{eq:05}
\end{equation}
where $z$ is the height from the midplane of the disk and $H_\mathrm{g}$ is the scale height of the disk given by
\begin{equation}
H_\mathrm{g} \equiv \frac{\sqrt{2} c_\mathrm{s}}{\Omega _\mathrm{K}} = 4.7 \times 10^{-2} \left( \frac{r}{1 [\mathrm{AU}]} \right)^{\frac{5}{4}} [\mathrm{AU}] \mathrm{.} \label{eq:06}
\end{equation}
The symbol $\Omega _\mathrm{K}$ is Keplerian angular velocity. The symbol $c_\mathrm{s}$ is the sound velocity as
\begin{equation}
c_\mathrm{s} = \sqrt{\frac{k_\mathrm{B} T}{m_\mu }} = 0.99 \left( \frac{r}{1 [\mathrm{AU}]} \right)^{-\frac{1}{4}} [\mathrm{km} \, \mathrm{s}^{-1}] \mathrm{,} \label{eq:07}
\end{equation}
where $k_\mathrm{B}$ is the Boltzmann constant and $m_\mu (= 3.9 \times 10^{-24} {~} \mathrm{g})$ is the mass of gas molecular (the mean molecular weight is 2.34). The mean free path of gas molecules, $l_\mathrm{g}$, is given by $l_\mathrm{g} = 1.44$ cm at $r = 1$ AU (\cite{key-N86}). From equation (\ref{eq:06}), the protoplanetary disk is geometrically thin.

\subsection{Dust properties}

In this paper, dust grains are assumed to be compact and spherical with radius $s$. In the case with $s \le 3 l_\mathrm{g} / 2 = 2.2$ cm, the gas drag force is given by Epstein's law. In this case, at $r = 1$ AU and $z = 0$, the stopping time $t_\mathrm{stop}$ is given by
\begin{equation}
t_\mathrm{stop} = \frac{\rho _\mathrm{s}}{\rho _\mathrm{g} (z)} \frac{s}{c_\mathrm{s}} = 1.5 \times 10^{-3} \left( \frac{f_\mathrm{g}}{1} \right)^{-1} \left( \frac{\rho _\mathrm{s}}{3 [\mathrm{g \, cm^{-3}}]} \right) \left( \frac{s}{2.2 [\mathrm{cm}]} \right) [\mathrm{year}] \mathrm{,} \label{eq:08}
\end{equation}
where $\rho _\mathrm{s}$ is the internal density of the dust grain. In the case with $s \ge 3 l_\mathrm{g} / 2$, gas drag is given by Stokes' law and the stopping time is given as
\begin{equation}
t_\mathrm{stop} = \frac{2}{3} \frac{\rho _\mathrm{s}}{\rho _\mathrm{g} (z)} \frac{s^2}{l_\mathrm{g} c_\mathrm{s}} = 1.5 \times 10^{-3} \left( \frac{l_\mathrm{g} f_\mathrm{g}}{1.4 [\mathrm{cm}]} \right)^{-1} \left( \frac{\rho _\mathrm{s}}{3 [\mathrm{g \, cm^{-3}}]} \right) \left( \frac{s}{2.2 [\mathrm{cm}]} \right)^2 [\mathrm{year}] \mathrm{,} \label{eq:09}
\end{equation}
at $r = 1$ AU and $z = 0$. In this paper, we adopt the criterion $s \le 3 l_\mathrm{g} / 2$ for the validity of the Epstein regime while there are also previous studies that use $s \le 9 l_\mathrm{g} / 4$ (e.g., \cite{key-Y02}).

In this paper, we assumed that radial motion of dust grains can be neglected (\cite{key-N81}). The equation of motion of a dust grain in the vertical direction is given by
\begin{equation}
\frac{d v_z}{d t} = - \frac{v_z}{t_\mathrm{stop}} - {\Omega _\mathrm{K}}^2 z \mathrm{,} \label{eq:10}
\end{equation}
where $t$ is the time and $v_z$ is the vertical velocity of the dust grain. We assume that gas are not affected by dust grains and gas density is always given by equation (\ref{eq:05}). In this paper, we approximate vertical velocity of dust grains by the terminal velocity (\cite{key-N81}; \cite{key-N86}). With setting $d v_z / dt = 0$ in equation (\ref{eq:10}), we obtain a terminal velocity of a dust grain as
\begin{equation}
v_z (z) = - t_\mathrm{stop} {\Omega _\mathrm{K}}^2 z \mathrm{.} \label{eq:11}
\end{equation}

\subsection{The mixture of gas and dust grains}

In this paper, we use a single-fluid approximation for the azimuthal motion. The rotational velocity of a mixed fluid of gas and dust is given as
\begin{equation}
v_\phi = \left[ 1 - \frac{\rho _\mathrm{g}(z)}{\rho _\mathrm{g}(z) + \rho _\mathrm{d} (z)} \eta \right] v_\mathrm{K} \mathrm{.} \label{eq:12}
\end{equation}
In equation (\ref{eq:12}), $\rho _\mathrm{d}$ is the dust density, $v_\mathrm{K}$ is the circular Keplerian velocity, and
\begin{equation}
\eta = - \frac{1}{4} \frac{{H_\mathrm{g}}^2}{r^2} \frac{\partial \ln P}{\partial \ln r} \mathrm{,} \label{eq:13}
\end{equation}
where $P = {c_\mathrm{s}}^2 \rho _\mathrm{g}$ is the gas pressure. Using equations (\ref{eq:02}), (\ref{eq:05}), (\ref{eq:06}) and (\ref{eq:07}), we have
\begin{equation}
\eta = \frac{13}{16} \left( \frac{H_\mathrm{g}}{r} \right)^2 = 1.8 \times 10^{-3} \left( \frac{r}{1 [\mathrm{AU}]} \right)^{\frac{1}{2}} \ll 1 \mathrm{.} \label{eq:14}
\end{equation}

\subsection{Richardson number}

To discuss the possibility of the shear-driven turbulence, we calculate Richardson number that is the indicator of KHI (\cite{key-C61}). The Richardson number is given by
\begin{equation}
\mathrm{Ri} = - \frac{g_z}{\rho _\mathrm{g}(z) + \rho _\mathrm{d} (z)} \left[ \frac{\partial \rho _\mathrm{g}(z)}{\partial z} + \frac{\partial \rho _\mathrm{d} (z)}{\partial z} \right] \left( \frac{\partial v_\phi }{\partial z} \right)^{-2} \mathrm{,} \label{eq:15}
\end{equation}
where $g_z$ is the gravitational acceleration given by $g_z = {\Omega _\mathrm{K}}^2 z$ (\cite{key-S01}). Equation (\ref{eq:15}) is based on an assumption that the disk is composed of an incompressible one component fluid for simplicity. We use equation (\ref{eq:15}) as an indicator of KHI for simplicity. If $\mathrm{Ri}$ is smaller than the critical value $\mathrm{Ri}_\mathrm{c}$, KHI is expected to be induced in the protoplanetary disk, and it is expected that the laminar flow of the mixed fluid in the disk becomes turbulent. \citet{key-C61} showed that $\mathrm{Ri}_\mathrm{c} = 0.25$, but \citet{key-G05} and \citet{key-J06} showed that the inclusion of the Coriolis force yields a much higher critical Richardson numbers. In this paper, we use a value $\mathrm{Ri}_\mathrm{c} = 0.25$, but in the subsequent section we will discuss the case with $\mathrm{Ri}_\mathrm{c} = 0.8$ (\cite{key-J06}).

By seeing equations (\ref{eq:12}) and (\ref{eq:15}), $\mathrm{Ri}$ depends on the distribution of dust density. In this paper, we calculate Richardson number for the dust density given by numerical calculations in each time.

\subsection{The growth and sedimentation of dust grains}

The growth and sedimentation of dust grains is described by
\begin{eqnarray}
\frac{\partial}{\partial t} & & n (m, z) + \frac{\partial}{\partial z} [n (m, z) v_z (m, z)] \nonumber \\
&&  = - n (m, z) \int_0^\infty A(m, m', z) n (m', z) dm' \nonumber \\
&& {~~~~~} + \frac{1}{2} \int_0^m [A(m - m', m', z) n (m - m', z) n (m', z) dm'] \mathrm{,} \label{eq:16}
\end{eqnarray}
where $n (m, z) dm$ is the number density of the dust grains with mass between $m$ to $m + dm$ at the height $z$ and $A(m, m', z)$ is the coalescence rate for two dust grains with $m$ and $m'$ at $z$. The symbol $n (m, z) dz$ gives a mass function in $z$ to $z + dz$. The dust density at $z$, $\rho _\mathrm{d} (z)$, is given by
\begin{equation}
\rho _\mathrm{d} (z) = \int_0^\infty m n (m, z) dm \mathrm{.} \label{eq:17}
\end{equation}

As for velocity which induces dust-dust collisions, velocities generated during sedimentation and the thermal Brownian motion are considered. The relative velocities of two dust grains due to sedimentation $\Delta v_\mathrm{s}$ and to the thermal motion $\Delta v_\mathrm{B}$ is given by
\begin{equation}
\Delta v_\mathrm{s} = |v_z (m, z) - v_z (m', z)| \mathrm{,} \label{eq:18}
\end{equation}
\begin{equation}
\Delta v_\mathrm{B} = \sqrt{k_\mathrm{B} T} \sqrt{\frac{1}{m} + \frac{1}{m'}} \mathrm{,} \label{eq:19}
\end{equation}
respectively. We assume that the coalescence rate $A(m, m', z)$ is given by
\begin{equation}
A(m, m', z) = \pi (s + s')^2 (\Delta v_\mathrm{s} + \Delta v_\mathrm{B}) p_\mathrm{s} \mathrm{,} \label{eq:20}
\end{equation}
where $p_\mathrm{s}$ is the coalescence probability and $p_\mathrm{s}$ is assumed to be 1 as in \citet{key-N81}. For simplicity, we ignore the bouncing barrier (\cite{key-Z10}) and the electrostatic barrier (\cite{key-O09a}).

Integrating equation (\ref{eq:16}) with respect to $m$, we have
\begin{equation}
\frac{\partial}{\partial t} \int_0^\infty m n (m, z) dm + \frac{\partial}{\partial z} \int_0^\infty m n (m, z) v_z (m, z) dm = 0 \mathrm{.} \label{eq:21}
\end{equation}
Using equations (\ref{eq:17}) and (\ref{eq:21}), we have
\begin{equation}
\frac{\partial}{\partial t} \rho _\mathrm{d} (z) + \frac{\partial}{\partial z} [\rho _\mathrm{d} (z) \bar{v}_z (z)] = 0 \mathrm{,} \label{eq:22}
\end{equation}
which is the continuity equation for dust grains treated as fluid, where the mean sedimentation velocity of dust fluid at $z$ is given by
\begin{equation}
\bar{v}_z (z) = \frac{\int_0^\infty v_z (z) \, m n (m, z) dm}{\int_0^\infty m n (m, z) dm} \mathrm{.} \label{eq:23}
\end{equation}

\section{Sedimentation of dust grains without growth}

In this section, in order to clarify the effect of dust growth, we first consider the case at $r = 1$ AU without the growth.

\subsection{The initial condition}

In this paper, dust density $\rho _\mathrm{d} (z)$ is assumed to be symmetric to the midplane, so we examine only the region of $z \ge 0$. As initial condition, the initial density of dust is assumed to be
\begin{equation}
\rho _\mathrm{d} (z) = \frac{\Sigma _\mathrm{d}}{\sqrt{\pi } H_\mathrm{d}} \exp \left[ - \left( \frac{z}{H_\mathrm{d}} \right)^2 \right] \mathrm{,} \label{eq:24}
\end{equation}
where $H_\mathrm{d}$ is time dependent scale height of dust density profile with $H_\mathrm{d} = H_\mathrm{g}$ at $t = 0$.

In this paper, we assume the same initial functional form for $\rho _\mathrm{g} (z)$ and $\rho _\mathrm{d} (z)$, and we neglect the dependence of $v_\mathrm{K}$ on $z$. Note that $\mathrm{Ri} = \infty $ at $t = 0$ from equation (\ref{eq:15}), i.e., initial state is stable against KHI. This is because the rotational velocity of the mixed fluid of gas and dust grains $v_\phi$ at $t = 0$ is independent of $z$ (cf. equation (\ref{eq:12})), i.e., $\partial v_\phi / \partial z = 0$ in equation (\ref{eq:15}).

\subsection{Sedimentation of dust grains with single size}

First, we consider the sedimentation of dust grains with a single size. We assume that all dust grains are small enough and that the drag force is given by Epstein's law. The characteristic time scale for sedimentation of dust grains is given by
\begin{equation}
t_\mathrm{sed} \equiv \frac{z}{|v_z (z)|} = \frac{1}{t_\mathrm{stop} {\Omega _\mathrm{K}}^2} \mathrm{.} \label{eq:25}
\end{equation}
From equations (\ref{eq:08}) and (\ref{eq:25}), it is shown that the characteristic time scale of sedimentation is much larger than the Keplerian period $t_\mathrm{sed} \Omega _\mathrm{K} \gg 1$ in the case when the stopping time is much smaller than the Keplerian period $t_\mathrm{stop} \Omega _\mathrm{K} \ll 1$. In such a case with a single size without dust growth, profile of dust density evolves in a self-similar manner (\cite{key-G04}). Here, during sedimentation of dust, the distribution of gas density assumed to be constant remain as equation (\ref{eq:05}). When the time evolution of dust density proceeds in a self-similar manner, dust density profile in $t > 0$ is also given by equation (\ref{eq:24}) with temporally decreasing $H_\mathrm{d} (t)$ with $0 < H_\mathrm{d} (t) \le H_\mathrm{g}$. Using equations (\ref{eq:05}), (\ref{eq:12}) and (\ref{eq:24}), we have
\begin{equation}
\frac{\partial v_\phi }{\partial z} = 2 \eta z v_\mathrm{K} \left( \frac{1}{{H_\mathrm{g}}^2} - \frac{1}{{H_\mathrm{d}}^2} \right) \frac{\rho _\mathrm{g}(z) \rho _\mathrm{d} (z)}{[\rho _\mathrm{g}(z) + \rho _\mathrm{d} (z)]^2} \mathrm{.} \label{eq:26}
\end{equation}
Substituting (\ref{eq:14}) and (\ref{eq:26}) into (\ref{eq:15}), an analytical formula for Richardson number can be derived.

\begin{figure}
  \begin{center}
    \FigureFile(74mm,74mm){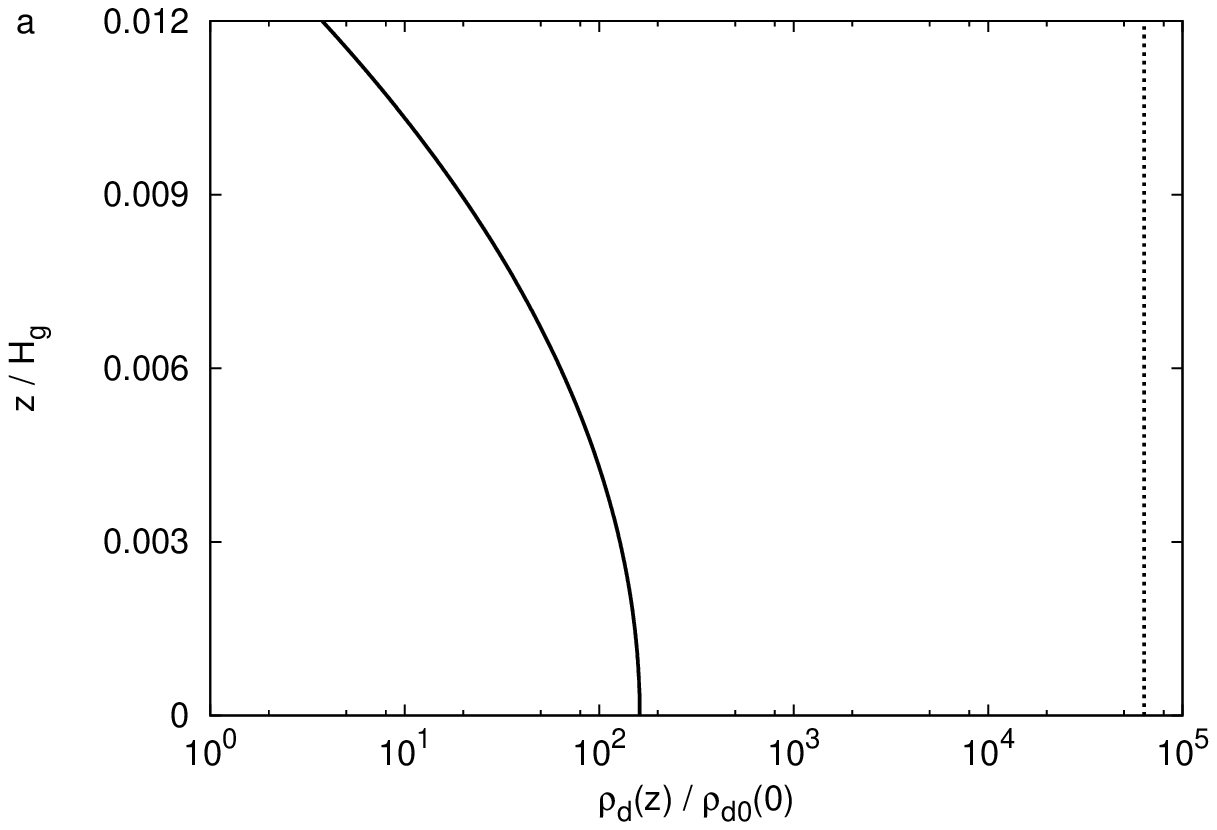}
    \FigureFile(74mm,74mm){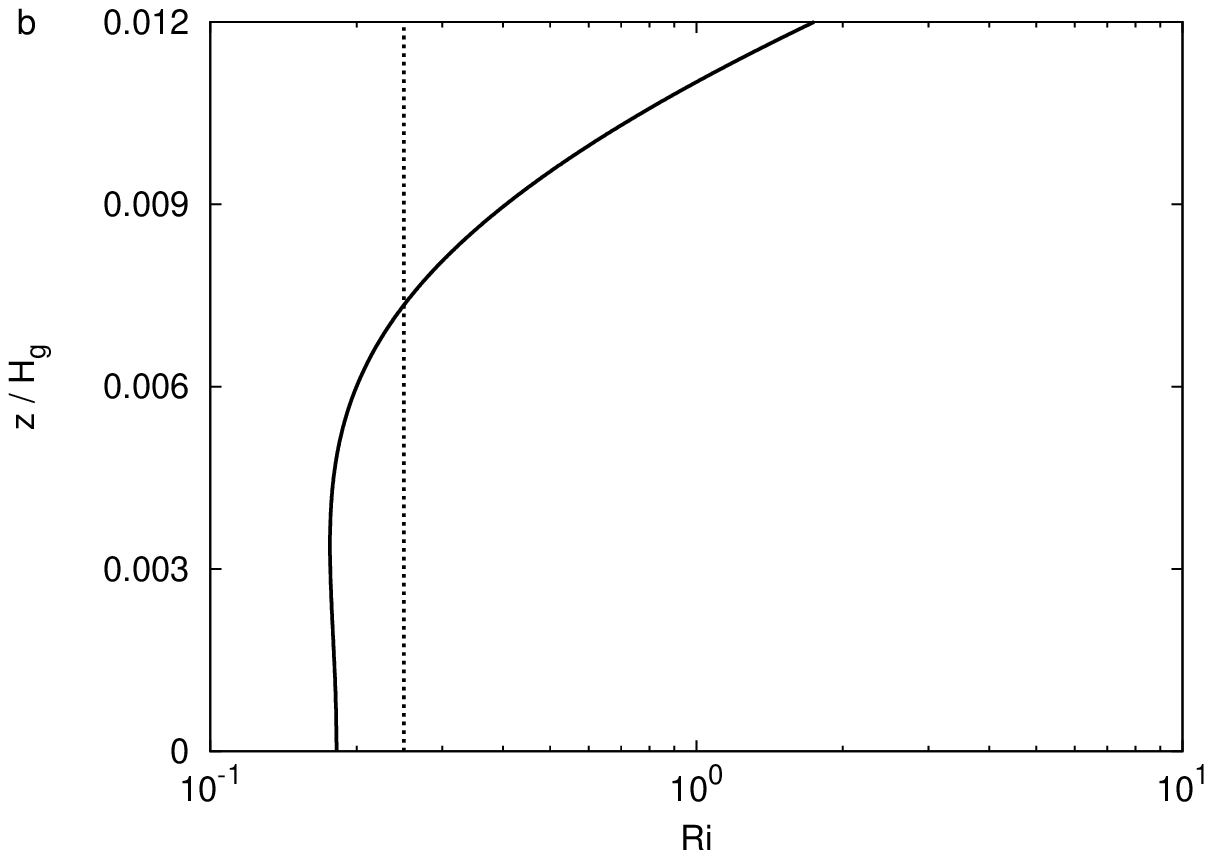}
  \end{center}
  \caption{Dust density and Richardson number at $r = 1$ AU for the case when dust grains have a single size with $f_\mathrm{d} = 1$. (a) The distribution of dust density (solid line). The abscissa shows the dust density $\rho _\mathrm{d} (z)$ in unit of $\rho _\mathrm{d0} (0)$. The ordinate shows $z$ coordinates in unit of $H_\mathrm{g}$. The critical density $\rho _\mathrm{c}$ is also drawn (dotted line). (b) The distribution of Richardson number (solid line). The abscissa shows Richardson number. The ordinate shows $z$ coordinates in unit of $H_\mathrm{g}$. The critical value $\mathrm{Ri}_\mathrm{c}$ is also drawn (dotted line).}\label{fig:01}
\end{figure}

Figure \ref{fig:01} shows results of distribution of dust density and Richardson number when $\rho _\mathrm{d} (0) \simeq 160 \rho _\mathrm{d0} (0)$ for the case with $f_\mathrm{d} = 1$, where $H_\mathrm{d}$ is $0.0062 H_\mathrm{g}$. We define the initial value of dust density at the midplane as $\rho _\mathrm{d0} (0) \equiv \Sigma _\mathrm{d} / (\sqrt{\pi } H_\mathrm{g})$. In Figure \ref{fig:01}, it is seen that dust density at the midplane is much smaller than the critical density $\rho _\mathrm{c} =6.3 \times 10^4 \rho _\mathrm{d0} (0)$, and that Richardson number is smaller than $\mathrm{Ri}_\mathrm{c} = 0.25$ around the midplane ($z / H_\mathrm{g} \lesssim 0.0075$). Thus, KHI is expected before GI in this case with $f_\mathrm{g} = f_\mathrm{d} = 1$ at $r = 1$ AU.

Assuming equilibrium condition for KHI, previous studies have shown that GI tends to occur if dust abundance in the protoplanetary disk is much larger than MMSN model (\cite{key-S98}; \cite{key-S01}). Now using non-equilibrium time dependent density profile during sedimentation, we consider the possibility for GI in the case with large $f_\mathrm{d}$. We seek the condition of dust abundance $f_\mathrm{d}$ by which GI occurs before the onset of KHI. In Figure \ref{fig:01} (a), it is seen that the distribution of dust density has a maximum value at the midplane. When $f_\mathrm{d}$ is larger than 1, the dotted line in Figure \ref{fig:01} (a) moves to left because the abscissa is inversely proportional to $\rho _\mathrm{d0} (0) \propto f_\mathrm{d}$. Dust density at the midplane is the first to reach the critical density for GI because time development of dust density proceeds in a self-similar manner. A characteristic $H_\mathrm{d}$ when dust density at midplane attains $\rho _\mathrm{c}$ can be obtained from equation (\ref{eq:24}). We define the characteristic $H_\mathrm{d}$ as $H_\mathrm{c}$. With substituting 0, $H_\mathrm{d}$ and $\rho _\mathrm{c}$ for $z$, $H_\mathrm{g}$ and $\rho _\mathrm{d} (z)$ in equation (\ref{eq:24}), respectively, $H_\mathrm{c}$ is given by
\begin{equation}
H_\mathrm{c} = \frac{\Sigma _\mathrm{d}}{\sqrt{\pi } \rho _\mathrm{c}} = 1.6 \times 10^{-5} f_\mathrm{d} H_\mathrm{g} \propto f_\mathrm{d} \mathrm{.} \label{eq:27}
\end{equation}
From Figure \ref{fig:01} (b), it is found that the distribution of Richardson number has a local minimum value at $z / H_\mathrm{g} = 3.5 \times 10^{-3}$. The height where distribution of Richardson number takes the local minimum value is defined as $z_\mathrm{c}$. At $z = z_\mathrm{c}$, from $\partial \mathrm{Ri} / \partial z |_{z = z_\mathrm{c}} = 0$, we find
\begin{equation}
\rho _\mathrm{g}(z_\mathrm{c}) = 2 \rho _\mathrm{d} (z_\mathrm{c}) \mathrm{,} \label{eq:28}
\end{equation}
with assuming $H_\mathrm{d} / H_\mathrm{g} \ll 1$. From equation (\ref{eq:28}), $z_\mathrm{c}$ is given as
\begin{equation}
z_\mathrm{c} = \left[ \ln \left( \frac{2 \Sigma _\mathrm{d}}{\Sigma _\mathrm{g}} \frac{H_\mathrm{g}}{H_\mathrm{d}} \right) \right] ^\frac{1}{2} H_\mathrm{g} \mathrm{.} \label{eq:29}
\end{equation}
Richardson number at $z = z_\mathrm{c}$ is given by
\begin{equation}
\mathrm{Ri} (z = z_\mathrm{c}) = \frac{27}{8} \left( \frac{H_\mathrm{d}}{\eta r} \right) ^2 \mathrm{,} \label{eq:30}
\end{equation}
with assuming $H_\mathrm{d} / H_\mathrm{g} \ll 1$. From equations (\ref{eq:27}), (\ref{eq:28}) and (\ref{eq:30}), the condition of dust abundance $f_\mathrm{d}$ that is necessary for GI to occur before KHI is derived as
\begin{equation}
f _\mathrm{d} = \left[ \frac{8 \pi }{27} \mathrm{Ri} (z = z_\mathrm{c}) \right] ^\frac{1}{2} \eta r \rho _\mathrm{c} \left( \frac{\Sigma _\mathrm{d}}{f_\mathrm{d}} \right) ^{-1} \ge 6.6 \times 10^2 \mathrm{,} \label{eq:31}
\end{equation}
at $r = 1$ AU and $f_\mathrm{g} = 1$. Note that $\Sigma _\mathrm{d} \propto f_\mathrm{d}$.

\begin{figure}
  \begin{center}
    \FigureFile(74mm,74mm){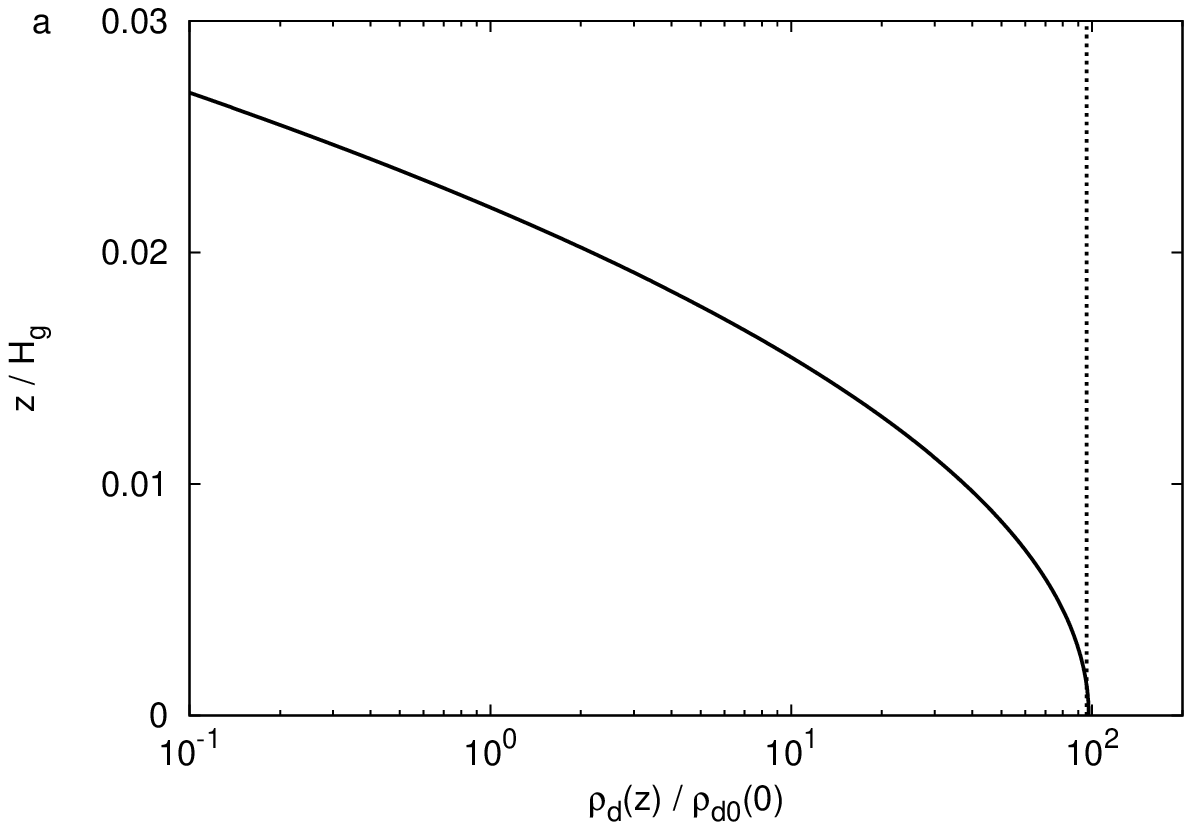}
    \FigureFile(74mm,74mm){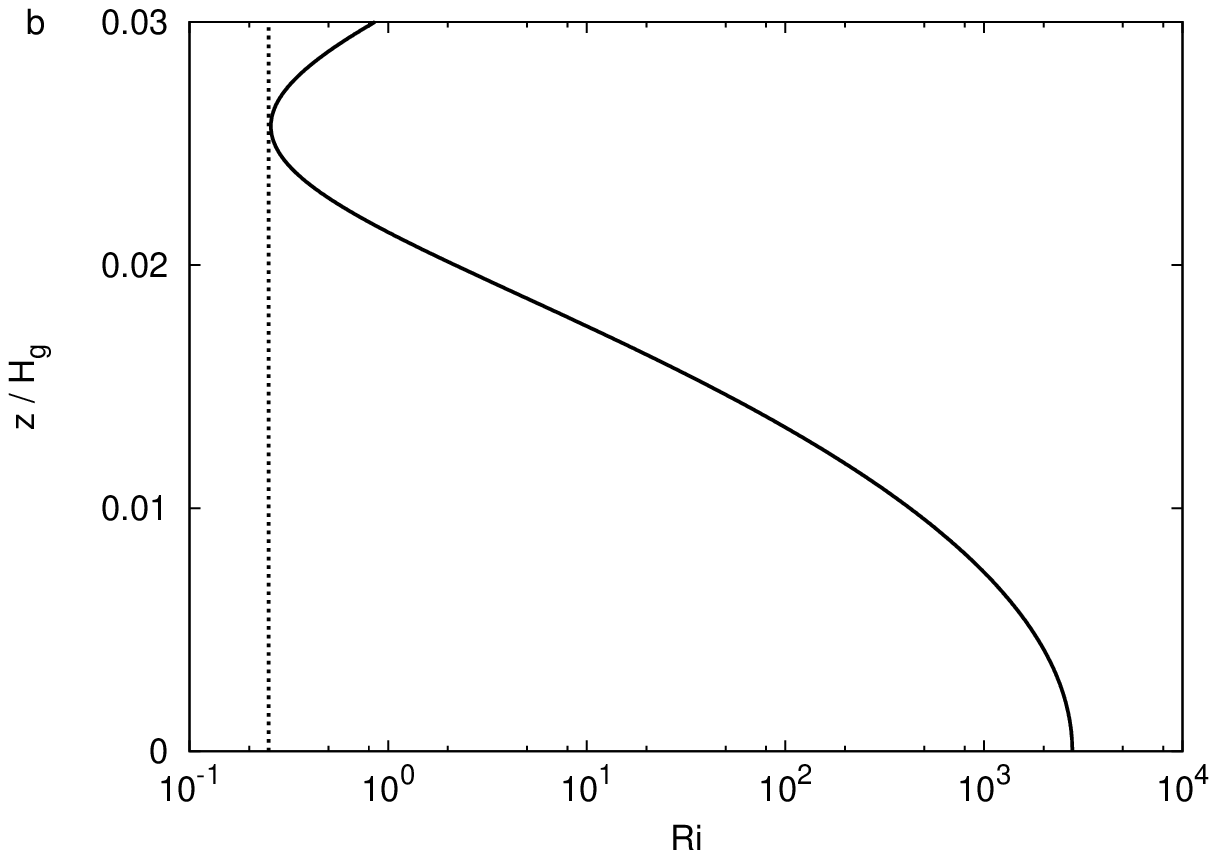}
  \end{center}
  \caption{Same as Figure \ref{fig:01}, but for the case with $f_\mathrm{d} = 6.6 \times 10^2$.}\label{fig:02}
\end{figure}

In Figure \ref{fig:02}, results for the case with $f_\mathrm{d} = 6.6 \times 10^2$ are shown. In Figure \ref{fig:02}, it is seen that dust density at the midplane indeed attains the critical density $\rho _\mathrm{c}$ and that Richardson number remains marginally larger than the critical value $\mathrm{Ri}_\mathrm{c}$. Thus, in the case when dust grains have a single size and don't grow at $r = 1$ AU with $f_\mathrm{g} = 1$, GI is expected to occur before KHI only if the dust surface density is much larger than the gas surface density.

This tendency is similar to the result of \citet{key-S98}, but the value of $f_\mathrm{d}$ in this paper ($f_\mathrm{d} = 6.6 \times 10^2$) is larger than the value in \citet{key-S98} ($f_\mathrm{d} = 16.8$). The value of $f_\mathrm{d}$ in \citet{key-S98} corresponds to the condition that the protoplanetary disk is in a quasi-equilibrium state for KHI. On the other hand, the value in this paper corresponds to the condition that GI occurs before KHI during sedimentation and that is more stringent. Therefore, the value of $f_\mathrm{d}$ in this paper is larger than the value in \citet{key-S98}.

\subsection{The effect of size distribution}

Next, we consider the sedimentation of dust grains with initial size distribution but without growth. The result given by equation (\ref{eq:31}) is independent of dust radius as long as the stopping time is given by Epstein's law. Now we consider the effect of initial size distribution of dust grains. Characteristic time scale for sedimentation of dust grains in equation (\ref{eq:25}) depends on size of dust grains. Thus, even in the same time, scale height of dust density profile for different size of dust grains is different. Therefore, it is expected that total dust density profile will change from initial Gaussian profile in the case when dust grains have a size distribution. We consider $N_\mathrm{d}$ kinds of dust grains with different radius. For simplicity, we consider the case when $s$ is given by integer multiples of the minimum radius of dust grains $s_0$ and when the maximum value $s_{\mathrm{max}}$ is given by $N_\mathrm{d} s_0$.

In order to derive the scale height $H_{\mathrm{d}, \, s}$ of dust density profile for dust grains with a radius $s$, we assume that $\rho _\mathrm{g} (z)$ is uniform for simplicity. The stopping time of dust grains $t_\mathrm{stop} = (\rho _\mathrm{s} / \rho _\mathrm{g}) (s / c_\mathrm{s})$ is proportional to $s$ and is constant with $z$. The vertical velocity of dust grains $v_z (z) \propto t_\mathrm{stop} z$ is proportional to $s$ and to $z$. From the time evolution of $z (t)$ with different radius of dust grains with $z = H_\mathrm{g}$ at $t = 0$, the formula for $H_{\mathrm{d}, \, s}$ can be derived as
\begin{equation}
\frac{H_{\mathrm{d}, \, s}}{H_\mathrm{g}} = \left( \frac{H_{\mathrm{d}, \, s_0}}{H_\mathrm{g}} \right)^{s / s_0} \mathrm{.} \label{eq:32}
\end{equation}
In equation (\ref{eq:32}), it is seen that $H_{\mathrm{d}, \, s}$ is determined only by $H_{\mathrm{d}, \, s_0}$ and $s / s_0$ instead of $s$ because radii of all dust grains are normalized by the smallest dust grains.

The distribution of total dust density $\rho _\mathrm{d} (z)$ is given as
\begin{equation}
\rho _\mathrm{d} (z) = \sum _{s / s_0 = 1}^{N_\mathrm{d}} \rho _\mathrm{d} (s, z) \mathrm{,} \label{eq:33}
\end{equation}
where $\rho _\mathrm{d} (s, z)$ is the density of dust grains with radius $s$ at $z$. For simplicity, initial size distribution is assumed to be a power law of dust radius,
\begin{equation}
n_\mathrm{d} (s, z) = n_\mathrm{d} (s_0, z) \left( \frac{s}{s_0} \right)^p \mathrm{,} \label{eq:34}
\end{equation}
where $n_\mathrm{d} (s, z)$ is the initial condition of the number density for radius $s$ at $z$ with a power index $p$. We assume $p = -3$ for simplicity. In the case with $p = -3$, $s^3 n_\mathrm{d} (s, z)$ is equal to ${s_0}^3 n_\mathrm{d} (s_0, z)$ from equation (\ref{eq:34}). Then, initial condition of $\rho _\mathrm{d} (s, z)$ is given by
\begin{equation}
\rho _\mathrm{d} (s, z) = \frac{4}{3} \pi \rho _\mathrm{s} s^3 n_\mathrm{d} (s, z) = \frac{4}{3} \pi \rho _\mathrm{s} {s_0}^3 n_\mathrm{d} (s_0, z) = \rho _\mathrm{d} (s_0, z) \mathrm{.} \label{eq:35}
\end{equation}
In the case with $p = -3$, equation (\ref{eq:35}) shows that initial density of dust with different radius is the same. In this case, $\rho _\mathrm{d} (s, z)$ is given by
\begin{equation}
\rho _\mathrm{d} (s, z) = \frac{1}{N_\mathrm{d}} \rho _\mathrm{d} (z) = \frac{1}{N_\mathrm{d}} \frac{\Sigma _\mathrm{d}}{\sqrt{\pi } H_{\mathrm{d}, \, s}} \exp \left[ - \left( \frac{z}{H_{\mathrm{d}, \, s}} \right)^2 \right] \mathrm{.} \label{eq:36}
\end{equation}
In the case without dust growth, $\rho _\mathrm{d} (s, z)$ with different $s$ evolves independently, with different $H_{\mathrm{d}, \, s}$. From equations (\ref{eq:32}), (\ref{eq:33}) and (\ref{eq:36}), we can derive analytical solutions of $\rho _\mathrm{d} (z)$. We assume $N_\mathrm{d} = 1000$.

\begin{figure}
  \begin{center}
    \FigureFile(74mm,74mm){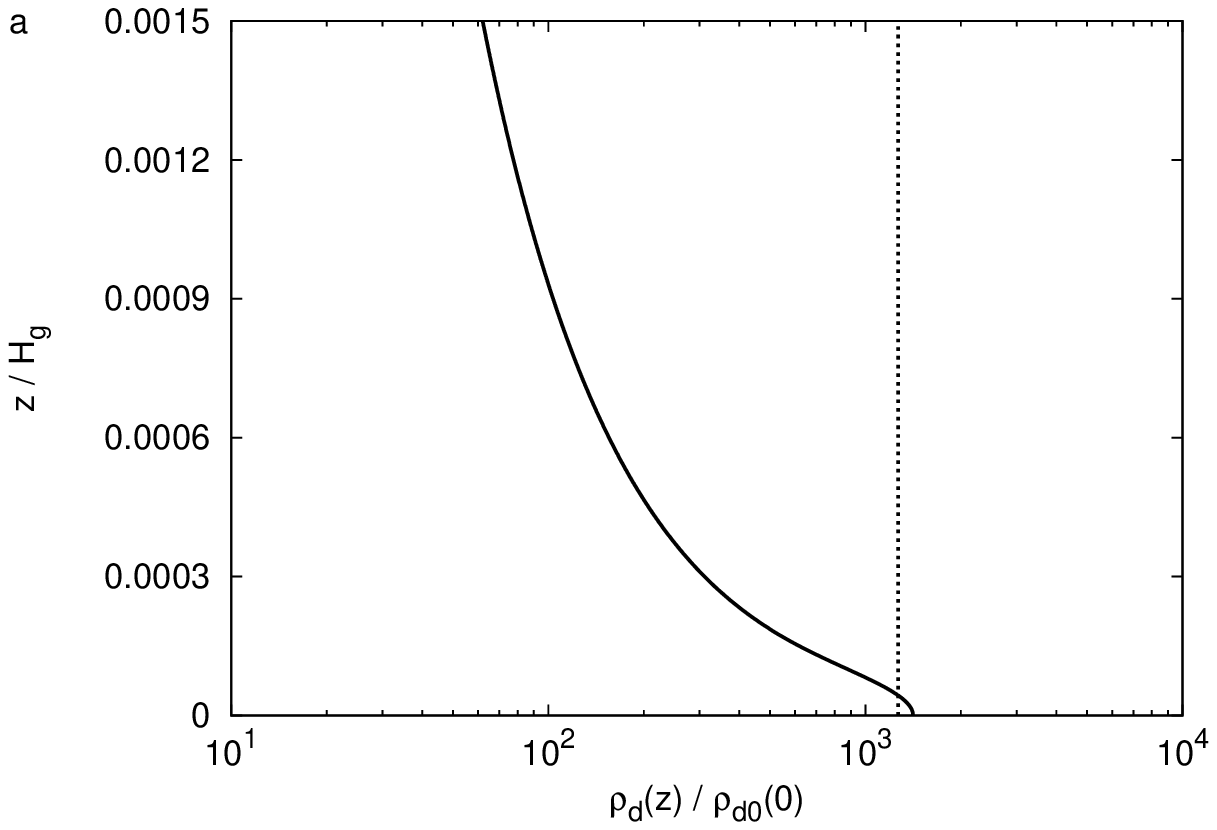}
    \FigureFile(74mm,74mm){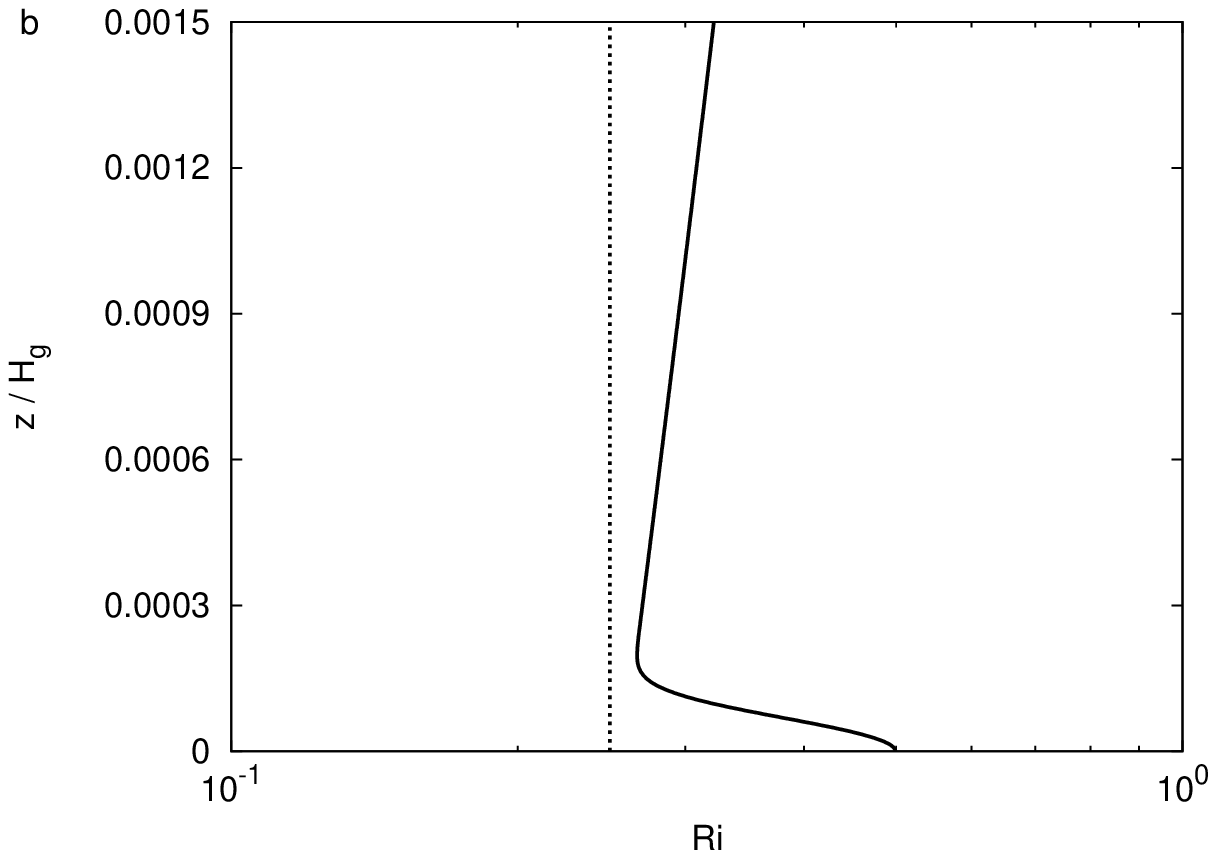}
  \end{center}
  \caption{Dust density and Richardson number at $r = 1$ AU for the case when dust grains have a size distribution with $f_\mathrm{d} = 50$. Lines, abscissas and ordinates show the same as ones of Figure \ref{fig:01}.}\label{fig:03}
\end{figure}

Figure \ref{fig:03} shows dust density and Richardson number for the case with $f_\mathrm{d} = 50$. In Figure \ref{fig:03}, it is seen that dust density at the midplane attains the critical density $\rho _\mathrm{c}$, and that Richardson number remains larger than the critical value. This demonstrates that dust fraction 50 times larger than MMSN model induces GI before KHI in the case without dust growth. Note that this dust fraction is about 10 times smaller than that in Figure \ref{fig:02}. We checked the dependence on $N_\mathrm{d}$ and found that this is the case with for $N_\mathrm{d} \gtrsim 10$. Thus, it can be suggested that KHI tends to be inhibited before GI if dust grains have size distribution even with the same $f_\mathrm{d}$. The reason is that the vertical gradient of dust density becomes smaller as a result of the continuous size distribution.

\begin{figure}
  \begin{center}
    \FigureFile(74mm,74mm){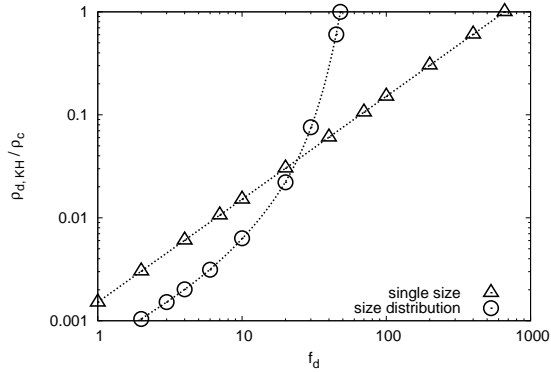}
  \end{center}
  \caption{The dust density in the midplane and in $r = 1$ AU at the onset of KHI for the case without size distribution of dust grains (triangles) and for the case with size distribution of dust grains (circles). The abscissa shows dust abundance $f_\mathrm{d}$, and the ordinate shows the dust density $\rho _\mathrm{d, KH}$ in unit of $\rho _\mathrm{c}$. The approximated curves are also drawn (dotted lines).}\label{fig:04}
\end{figure}

Figure \ref{fig:04} shows dust density in the midplane at the onset of KHI for the both cases without and with size distribution. The symbol $\rho _\mathrm{d, KH}$ is the dust density in the midplane at the onset of KHI. It is seen that dust density in the midplane at the onset of KHI increases with increasing the dust abundance in both cases with and without size distribution. Especially, it is seen that the condition $\rho _\mathrm{d, KH} = \rho _\mathrm{c}$ is attained by smaller dust abundance $f_\mathrm{d}$ in the case with size distribution $f_\mathrm{d} > 50$ than without size distribution $f_\mathrm{d} \gtrsim 700$.

Above results are based on $\mathrm{Ri}_\mathrm{c} = 0.25$ (\cite{key-C61}). However, \citet{key-J06} showed that $\mathrm{Ri}_\mathrm{c} = 0.8$. We also investigated the case with $\mathrm{Ri}_\mathrm{c} = 0.8$. In the case without size distribution, the required abundance of dust for GI is about $1.2 \times 10^3$. On the other hand, in the case with size distribution, the required abundance is smaller than 100. This abundance is much smaller than that for the case without size distribution, and this result is qualitatively same as that for the case with $\mathrm{Ri}_\mathrm{c} = 0.25$. Thus, below, we concentrate on the case with $\mathrm{Ri}_\mathrm{c} = 0.25$.

In the case without dust growth, results are summarized as follows:

\begin{enumerate}
\item In the case when the abundance of dust grains is given as MMSN model, KHI is expected to occur when dust density at the midplane is still much smaller than the critical density for GI.
\item GI tends to occur if the abundance of dust grains is larger.
\item If dust grains have the initial size distribution, the required abundance of dust for GI has the possibility to be smaller than that in the case without size distribution.
\end{enumerate}

Above results are based on the assumption that dust grains don't grow and that initial size distribution is assumed. However, in the actual protoplanetary disks, dust grains are expected to collide mutually and to grow. If dust grains grow, the size distribution and the largest radius of dust grains will change with time. In order to understand the actual condition of dust abundance for GI, size distribution which is consistent with dust growth during sedimentation is desirable. Next, we consider the case with dust growth.

\section{The case with dust growth}

\subsection{The initial condition and the numerical method}

We investigate sedimentation and growth of dust grains at $r = 1$ AU in the case with $f_\mathrm{g} = 1$. We solve equation (\ref{eq:16}) numerically. We assume that the internal density of dust grains $\rho _\mathrm{s}$ is 3.0 $\mathrm{g \, cm^{-3}}$. As initial condition, we assume that all dust grains have the initial size $s_0 = 1.0 \times 10^{-4}$ cm and initial density given by equation (\ref{eq:24}).

\begin{figure}
  \begin{center}
    \FigureFile(74mm,74mm){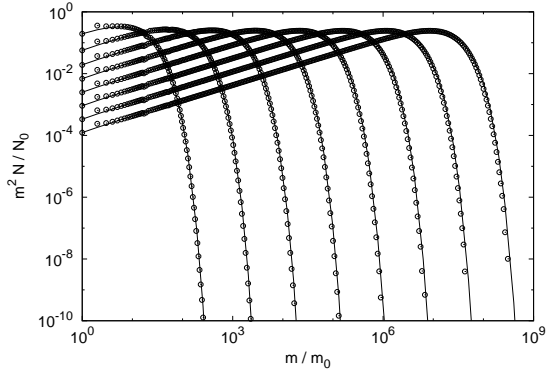}
  \end{center}
  \caption{Results of the test of our numerical method for growth of dust grains. Numerical results (circles) are compared with analytic solutions by \citet{key-T71} (solid lines).}\label{fig:05}
\end{figure}

In numerical calculations, TVD scheme (\cite{key-R86}) is applied to calculate sedimentation of dust grains and the method of WS89 (\cite{key-W89}) is applied for dust growth. The mass coordinates $m_i$ are logarithmically divided into 400 mass bins. Our numerical method is tested using the analytical solution (\cite{key-T71}). Figure \ref{fig:05} shows results with our numerical method for growth of dust grains. In Figure \ref{fig:05}, it is seen that our numerical method has satisfactory accuracy to calculate growth of dust grains. The $z$ coordinates $z_j$ are logarithmically divided into 106 spaced grids. The thickness of the nearest grid to the midplane is $1.4 \times 10^{-8} H_\mathrm{g}$. Using this spatial resolution, scale height at GI is sufficiently resolved.

\subsection{Possibilities of KHI in the early stage}

\begin{figure}
  \begin{center}
    \FigureFile(74mm,74mm){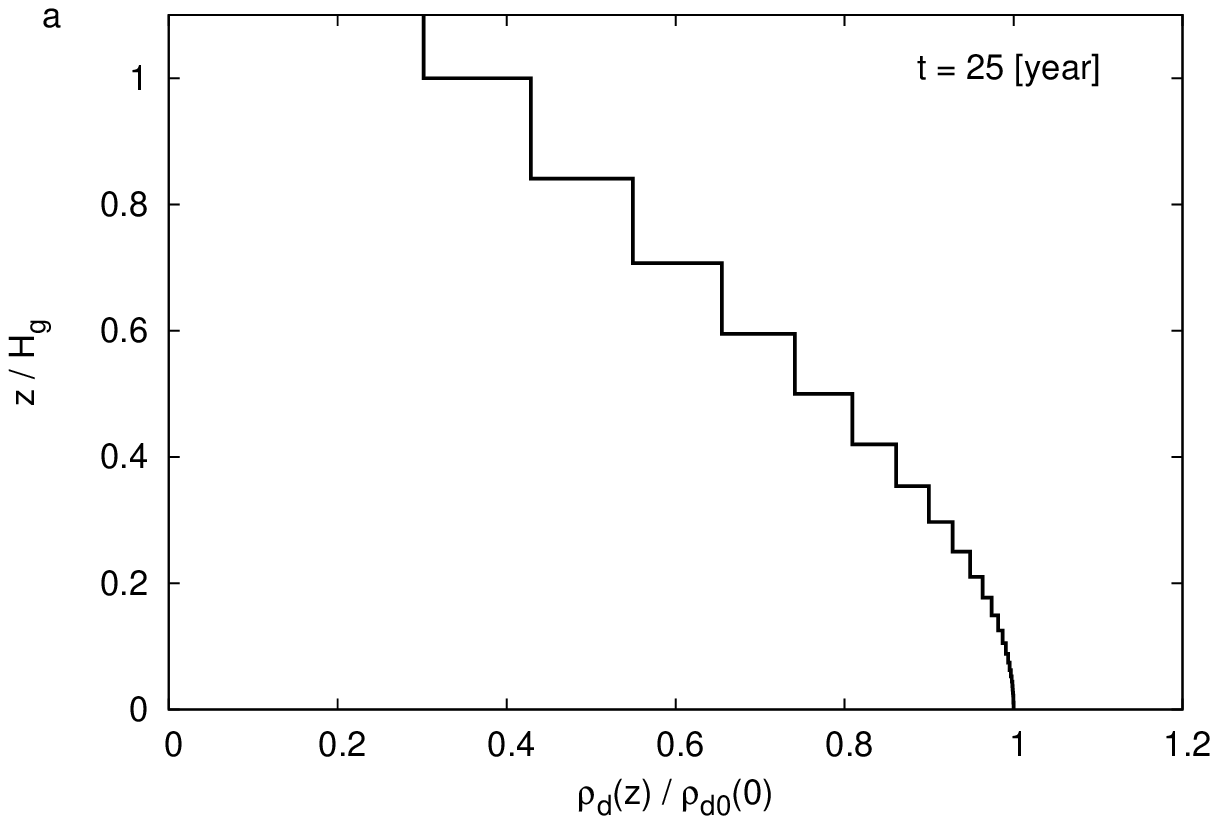}
    \FigureFile(74mm,74mm){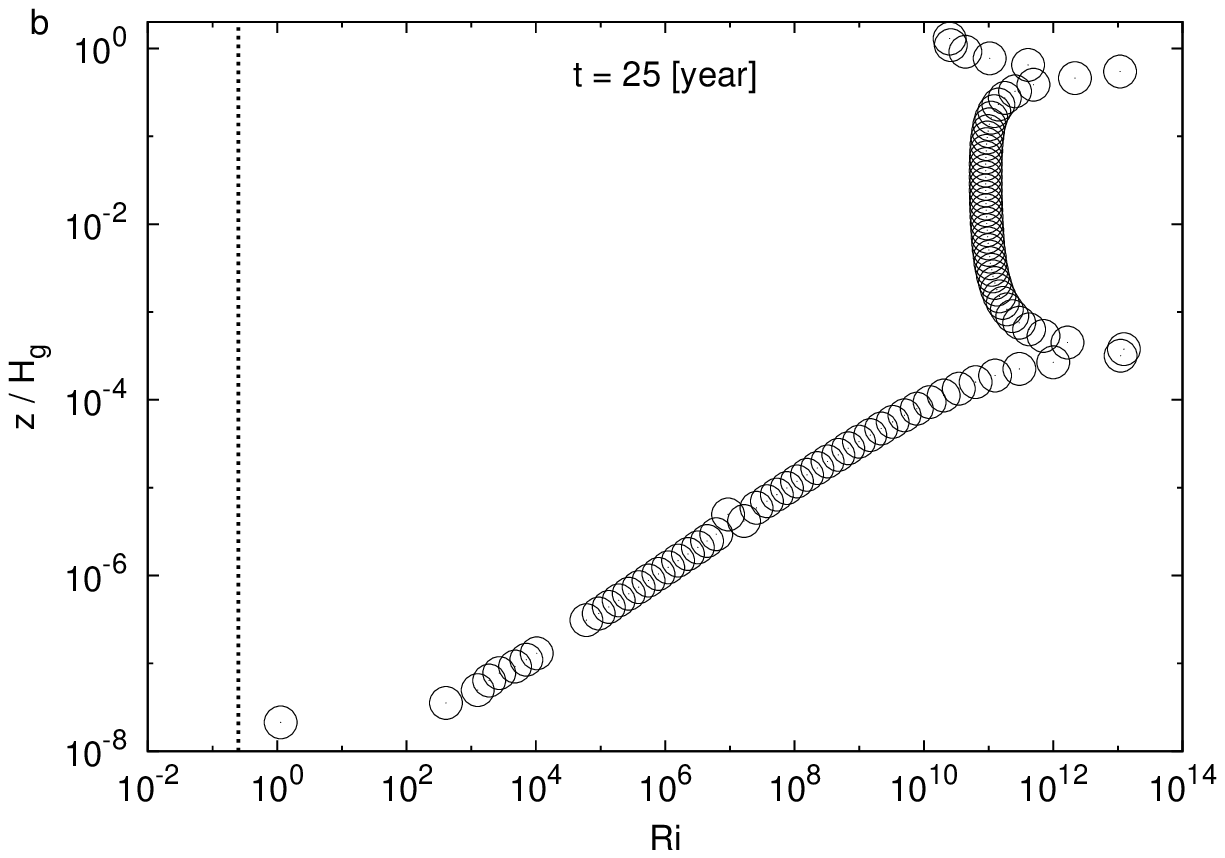}
  \end{center}
  \caption{(a) The distribution of dust density at $r = 1$ AU and at $t = 25$ year. The line, the abscissa and the ordinate show the same as ones of Figure \ref{fig:01} (a). (b) The distribution of Richardson number at $t = 25$ year. The abscissa shows Richardson number. The ordinate shows $z$ coordinates in unit of $H_\mathrm{g}$. The critical value $\mathrm{Ri}_\mathrm{c}$ is also drawn (dotted line).}\label{fig:06}
\end{figure}

Figure \ref{fig:06} shows the distribution of dust density and Richardson number at $t = 25$ year. In Figure \ref{fig:06}, although the distribution of dust density changes little from the initial state in this short period, it is seen that Richardson number becomes small enough in small $z$ especially $z / H_\mathrm{g} \lesssim 10^{-4} \ll 1$. This rapid decline of $\mathrm{Ri}$ did not come out in the case without growth of dust grains in \S 3. We suppose that growth of dust grains is the origin of this rapid decline in Richardson number at $z / H_\mathrm{g} \ll 1$ and at $t / t_\mathrm{sed} \ll 1$.

Equations (\ref{eq:12}) and (\ref{eq:15}) show that Richardson number is a function of the gradient of dust density $\partial \rho _\mathrm{d} (z) / \partial z$, so we now think about $\partial \rho _\mathrm{d} (z) / \partial z$. First, we think about the case without growth of dust grains to evince the effect of dust growth. We think about the case when all dust grains have a single size as initial condition because we assume that all dust grains have an initial radius $s_0$ as initial condition for the calculation in the case with dust growth. In the case without growth and size distribution of dust grains, the typical radius of dust grains $\bar{s} (z)$, which is defined as
\begin{equation}
\bar{s}(z) \equiv \frac{\int_0^\infty s \, m n (m, z) dm}{\int_0^\infty m n (m, z) dm} \mathrm{,} \label{eq:37}
\end{equation}
is equal to $s_0$ and is independent on $z$. \citet{key-G04} shows that time dependence of dust density proceeds in the self-similar manner in the case without growth of dust grains. In this case, the gradient of dust density is given by
\begin{equation}
\frac{\partial \rho _\mathrm{d} (z)}{\partial z} = - \frac{2 \Sigma _\mathrm{d} z}{\sqrt{\pi } {H_\mathrm{d}}^3} \propto z \, \to \, 0 {~~~} (z \to 0) \mathrm{.} \label{eq:38}
\end{equation}
In order to compare with the case with growth of dust grains, we think about the sedimentation velocity for the gravity center of the dust system at $z$, $\bar{v}_z (z)$, which is derived from equations (\ref{eq:11}), (\ref{eq:23}) and (\ref{eq:37}). At $z / H_\mathrm{g} \ll 1$ and at $t / t_\mathrm{sed} \ll 1$, $\bar{v}_z (z)$ is given by
\begin{equation}
\bar{v}_z (z) \equiv - \frac{\rho _\mathrm{s}}{\rho _\mathrm{g} (z)} \frac{\bar{s}(z)}{c_\mathrm{s}} {\Omega _\mathrm{K}}^2 z \mathrm{.} \label{eq:39}
\end{equation}
In the case without growth of dust grains, $\bar{v}_z (z) \propto z$ at $z / H_\mathrm{g} \ll 1$ because $\bar{s} (z) = s_0$.

Second, we think about the case with growth of dust grains. In this case, the typical radius of dust grains $\bar{s} (z)$ is the linear function of $z$ at $z / H_\mathrm{g} \ll 1$ and at $t / t_\mathrm{sed} \ll 1$ owing to collisions due to sedimentation (see Appendix for the reason). From equation (\ref{eq:39}), $\bar{v}_z (z)$ has a second-order term of $z$ because the typical radius of dust grains is the linear function of $z$. From comparing the functional form of $\bar{v}_z (z)$ with that in the case without dust growth, at $z / H_\mathrm{g} \ll 1$ and at $t / t_\mathrm{sed} \ll 1$, it is suggested that the gradient of dust density is given by
\begin{equation}
\frac{\partial \rho _\mathrm{d} (z)}{\partial z} = \delta _1 z + \delta _2 \to \delta _2 {~~~~} (z \to 0) \mathrm{,} \label{eq:40}
\end{equation}
where $\delta _1 (< 0)$ and $\delta _2 (> 0)$ are appropriate values.

\begin{figure}
  \begin{center}
    \FigureFile(74mm,74mm){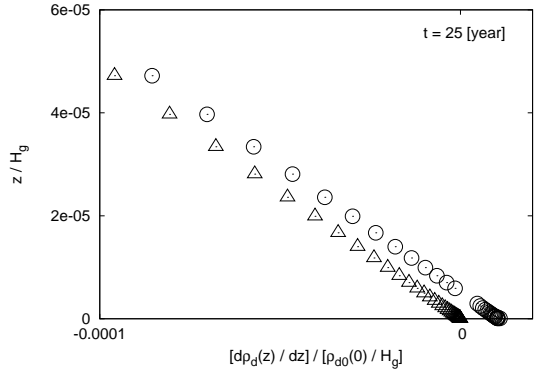}
  \end{center}
  \caption{The distribution of the gradient of dust density in $r = 1$ AU and $z / H_\mathrm{g} \ll 1$ at $t = 25$ year. The abscissa shows the gradient of dust density $\partial \rho _\mathrm{d} (z) / \partial z$ in unit of $\rho _\mathrm{d0} (0) / H_\mathrm{g}$. The ordinate shows $z$ coordinates in unit of $H_\mathrm{g}$. Circles show the case with dust growth and triangles show the one without dust growth.}\label{fig:07}
\end{figure}

\begin{figure}
  \begin{center}
    \FigureFile(74mm,74mm){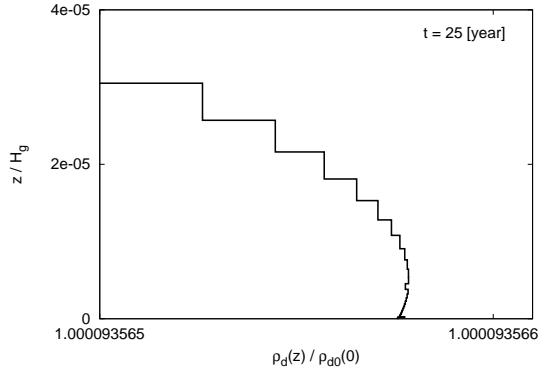}
  \end{center}
  \caption{The distribution of dust density in $r = 1$ AU and $z / H_\mathrm{g} \ll 1$ at $t = 25$ year in the case with growth of dust grains. The line, the abscissa and the ordinate show the same as ones of Figure \ref{fig:01} (a).}\label{fig:08}
\end{figure}

\begin{figure}
  \begin{center}
    \FigureFile(74mm,74mm){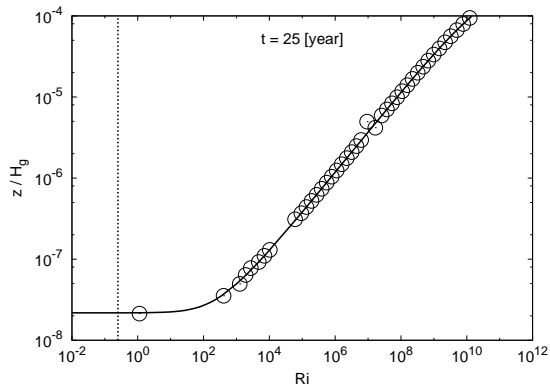}
  \end{center}
  \caption{The distribution of Richardson number in $r = 1$ AU and $z / H_\mathrm{g} \ll 1$ at $t = 25$ year in the case with growth of dust grains. Circles, the dotted line, the abscissa and the ordinate show the same as ones of Figure \ref{fig:06} (b). The solid line is given by equation (\ref{eq:41}).}\label{fig:09}
\end{figure}

Figure \ref{fig:07} shows the distribution of the gradient of the dust density at $z / H_\mathrm{g} \ll 1$ and at $t = 25$ year for cases with and without dust growth. From Figure \ref{fig:07}, it is confirmed that the distribution of the gradient of dust density is approximated by equation (\ref{eq:40}) with $\delta _2 > 0$ in the case with dust growth. From Figure \ref{fig:07}, we can approximate $\partial \rho _\mathrm{d} (z) / \partial z \simeq [ - 2.0 (z / H_\mathrm{g}) + 1.0 \times 10^{-5}] [\rho _\mathrm{d0} (0) / H_\mathrm{g}]$. This indicates that the distribution of dust density has a local maximum value at $z / H_\mathrm{g} \sim 5 \times 10^{-6}$ in the case with growth of dust grains. This can be confirmed by seeing Figure \ref{fig:08} that shows the distribution of dust density in $z / H_\mathrm{g} \ll 1$ and at $t = 25$ year in the case with dust growth. By approximating $\rho _\mathrm{g} (z)$ as $\rho _\mathrm{g} (0)$ and $\rho _\mathrm{d} (z)$ as $\rho _\mathrm{d0} (0)$, the Richardson number for $z / H_\mathrm{g} \ll 1$ is given by
\begin{eqnarray}
\mathrm{Ri} = \left( \frac{16}{13} \right)^2 & & \left( \frac{r}{H_\mathrm{g}} \right)^2 \left( \frac{\Sigma _\mathrm{g}}{\Sigma _\mathrm{d}} \right)^{-2} \left( \frac{\Sigma _\mathrm{g}}{\Sigma _\mathrm{d}} + 1 \right)^3 \nonumber \\
&& \times \frac{z}{H_\mathrm{g}} \left[ \frac{2 \Sigma _\mathrm{g}}{\Sigma _\mathrm{d}} \frac{z}{H_\mathrm{g}} - \frac{\partial \rho _\mathrm{d} (z)}{\partial z} \frac{H_\mathrm{g}}{\rho _\mathrm{d0} (0)} \right]  \left[ \frac{2 z}{H_\mathrm{g}} + \frac{\partial \rho _\mathrm{d} (z)}{\partial z} \frac{H_\mathrm{g}}{\rho _\mathrm{d0} (0)} \right]^{-2} \mathrm{.} \label{eq:41}
\end{eqnarray}
Figure \ref{fig:09} shows the distribution of Richardson number at $z / H_\mathrm{g} \ll 1$ and at $t = 25$ year in the case with growth of dust grains. In Figure \ref{fig:09}, it is seen that numerical solutions of Richardson number is sufficiently close to the approximate equation (\ref{eq:41}). In Figure \ref{fig:09}, it is seen that Richardson number drawn by the solid line is smaller than the critical value $\mathrm{Ri}_\mathrm{c}$ around the midplane at $t = 25$ year.

However, it is doubtful whether KHI occurs. In Figure \ref{fig:08}, it is seen that the distribution of dust density has a local maximum value at $z \neq 0$. Under the local maximum point, the Rayleigh-Taylor instability (RTI) is suspected to occur because the distribution of the gradient of dust density becomes positive (\cite{key-C61}). If RTI occurs, the distribution of dust density in this region is expected to be adjusted as to be constant (\cite{key-S01}). If it is assumed that the growth rate of RTI is larger than that of KHI, and that the distribution of dust density becomes constant around the midplane, the gradient of dust density becomes zero near the midplane. In this case with assuming that gas density is given by equation (\ref{eq:05}), Richardson number near the midplane is given by
\begin{eqnarray}
\mathrm{Ri} = \frac{128}{169} \left( \frac{r}{H_\mathrm{g}} \right)^2 \frac{(\Sigma _\mathrm{g} + \Sigma _\mathrm{d})^3}{\Sigma _\mathrm{g} {\Sigma _\mathrm{d}}^2} \sim 2 \times 10^7 \gg 0.25 = \mathrm{Ri}_\mathrm{c} \mathrm{.} \label{eq:42}
\end{eqnarray}
This indicates that KHI does not occur. Thus, below, the possibility of KHI near the midplane in the early phase as indicated in Figure \ref{fig:06} (b) is not considered further.

\subsection{The dust density at the onset of KHI for the case with dust growth}

\begin{figure}
  \begin{center}
    \FigureFile(74mm,74mm){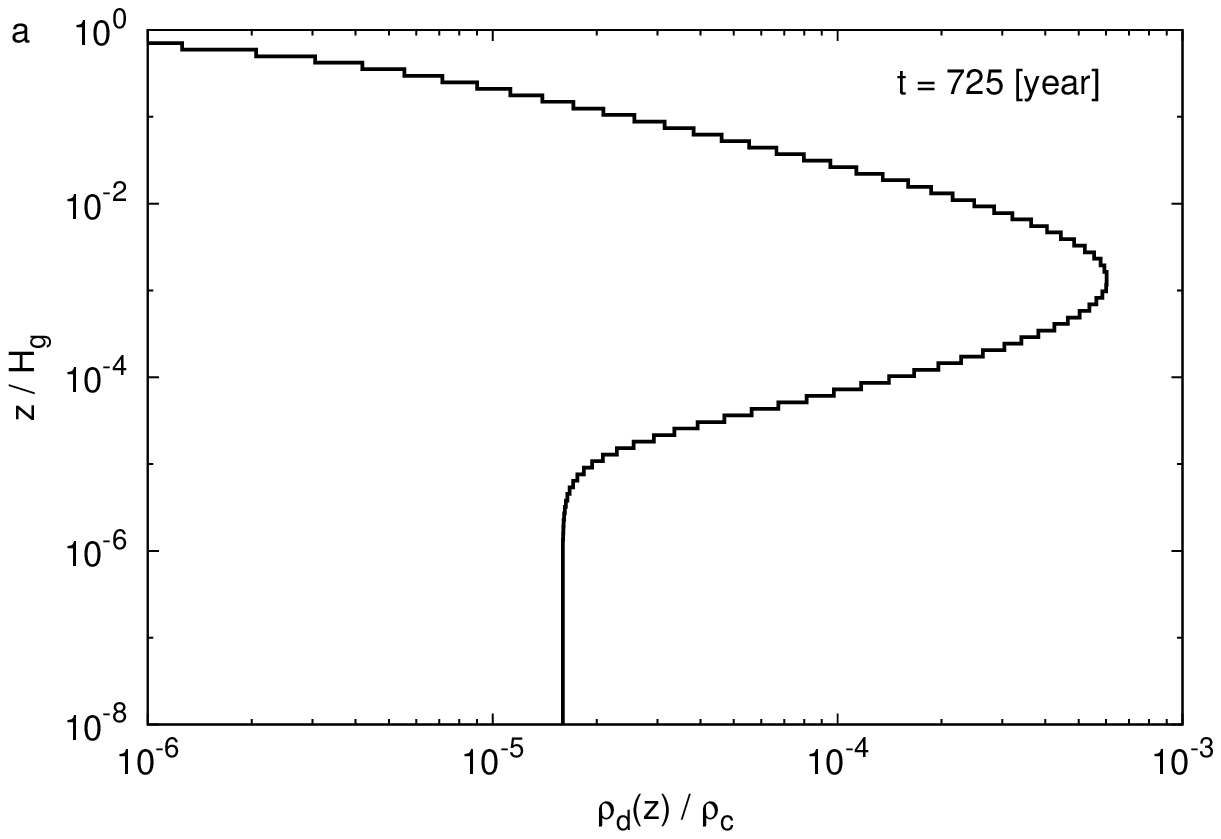}
    \FigureFile(74mm,74mm){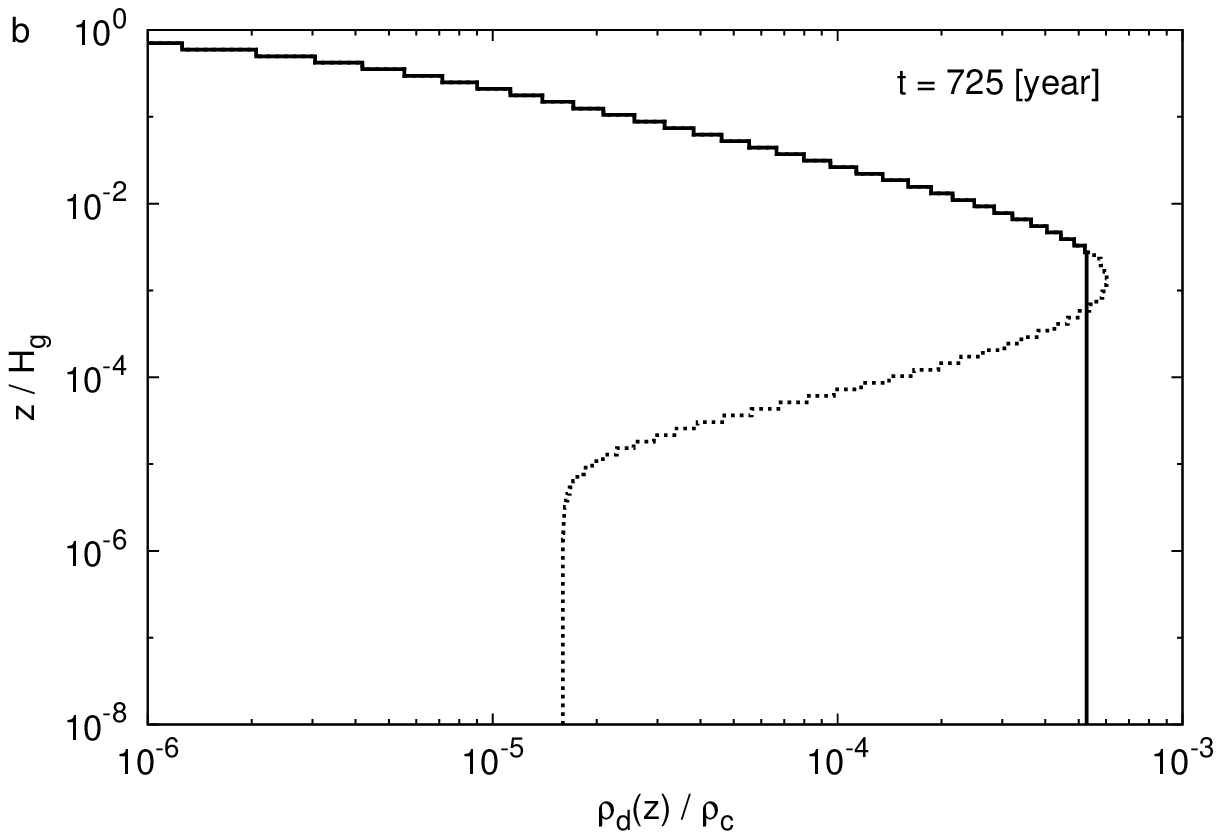}
  \end{center}
  \caption{Dust density in $r = 1$ AU at the onset of KHI for the case with dust growth and $f_\mathrm{d} = 1$. (a) The dust density obtained by numerical calculations. (b) The dust density adjusted as to be constant (solid line) and the dust density not adjusted (dotted line). The dust density not adjusted is equal to the one obtained by numerical calculations. For both of Figure (a) and (b), the abscissas show the dust density $\rho _\mathrm{d} (z)$ in unit of $\rho _\mathrm{c}$, and the ordinates show $z$ coordinates in unit of $H_\mathrm{g}$.}\label{fig:10}
\end{figure}

Figure \ref{fig:10} shows dust density distribution at the onset of KHI in the case with dust growth with $f_\mathrm{d} = 1$. In Figure \ref{fig:10} (a), distribution of dust density has a local maximum value at $z = z_\mathrm{RT} \sim 10^{-3} H_\mathrm{g}$, and there is the region where the gradient of dust density becomes positive. As discussed in \S 4.2, in this case, the distribution of dust density is expected to be adjusted to be constant by RTI in the region $z \lesssim z_\mathrm{RT}$. Assuming this RTI, we modify the distribution of dust density as Figure \ref{fig:10} (b) with mass conservation. Note that this treatment for RTI is crude and more accurate treatment should be addressed in the future. Hereafter, same modification is always applied for the density near the midplane.

\begin{figure}
  \begin{center}
    \FigureFile(74mm,74mm){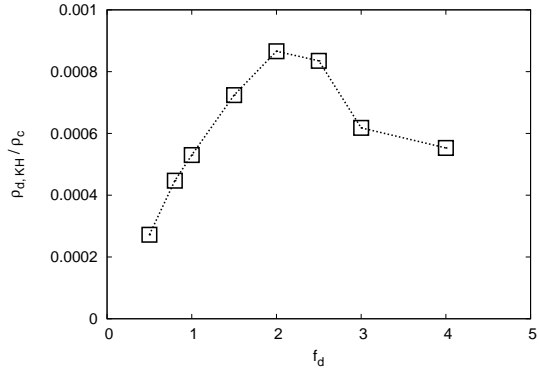}
  \end{center}
  \caption{The modified dust density in $r = 1$ AU at the onset of KHI for the case with dust growth. The abscissa and the ordinate show the same as ones of Figure \ref{fig:04}.}\label{fig:11}
\end{figure}

\begin{figure}
  \begin{center}
    \FigureFile(74mm,74mm){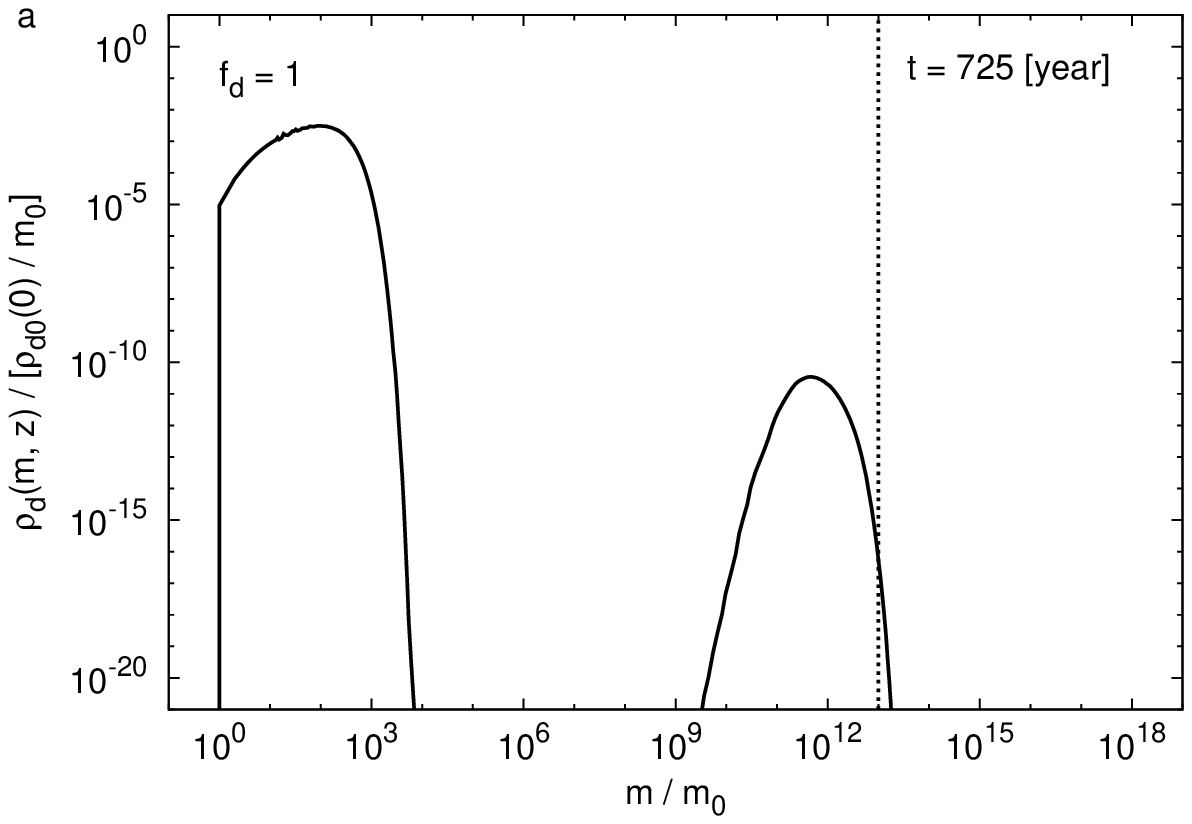}
    \FigureFile(74mm,74mm){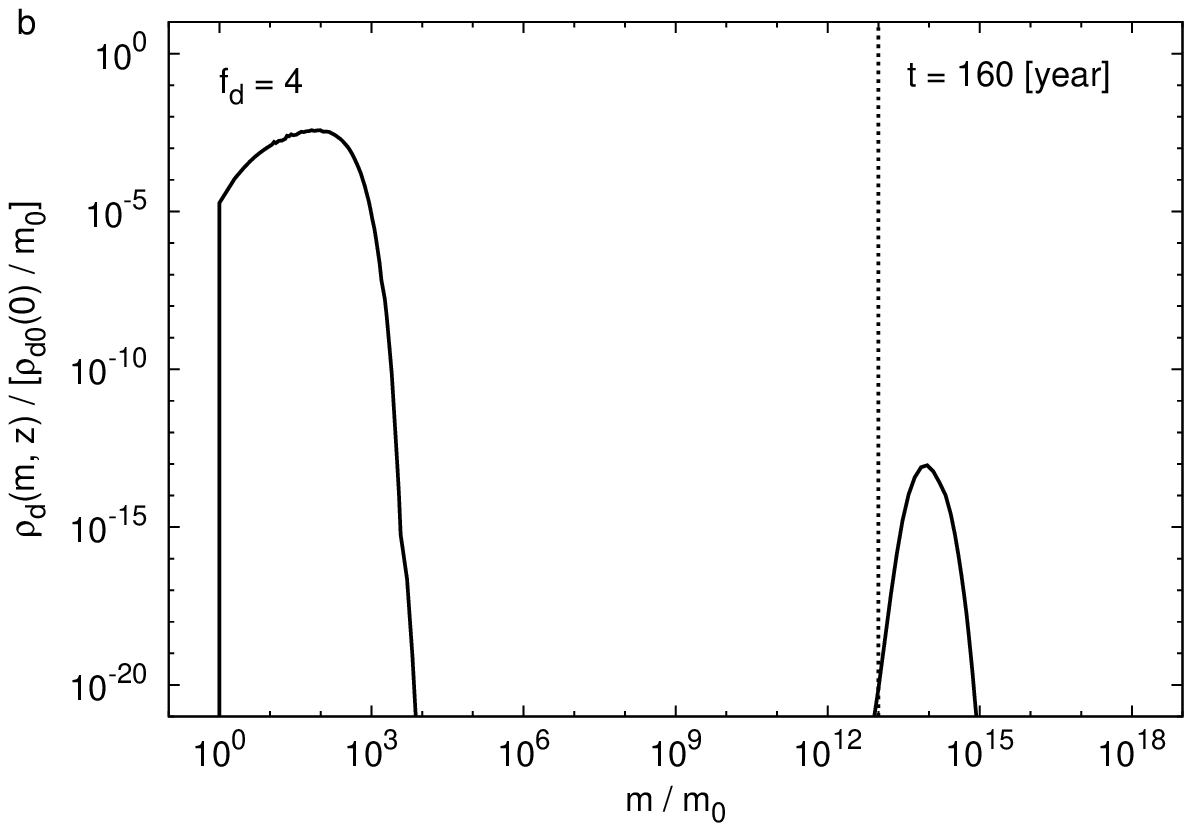}
  \end{center}
  \caption{(a) The mass function in the case with $f_\mathrm{d} = 1$ at $z = 1.5 \times 10^{-3} H_\mathrm{g}$ and $r = 1$ AU where the distributions of dust density takes a local maximum value at the onset of KHI. (b) The mass function in the case with $f_\mathrm{d} = 4$ at $z = 3.8 \times 10^{-4} H_\mathrm{g}$ and $r = 1$ AU where the distributions of dust density takes a local maximum value at the onset of KHI. In both of diagrams (a) and (b),  The abscissa shows the dust mass $m$ in unit of $m_0$. The ordinate shows the mass function $m n (m, z)$ in unit of $\rho _\mathrm{d0} (0) / m_0$. Dotted lines show the border line, where the stopping time changes from Epstein's law to Stokes' law.}\label{fig:12}
\end{figure}

Figure \ref{fig:11} shows the modified dust density profile at the onset of KHI. Figure \ref{fig:11} corresponds to Figure \ref{fig:04} with dust growth. Since the case with large $f_\mathrm{d}$ is numerically expensive, results only for the case with $f_\mathrm{d} \le 4$ are plotted. In Figure \ref{fig:11}, it is seen that dust density at KHI, $\rho _\mathrm{d, KH}$, increases with increasing dust abundance for the case with $f_\mathrm{d} < 2$. On the other hand, in the case with $f_\mathrm{d} > 2$, it is seen that dust density at KHI decreases with increasing dust abundance. This tendency is qualitatively different from Figure \ref{fig:04}. In Figure \ref{fig:11}, our results show that dust density in the midplane at the onset of KHI for the case with $f_\mathrm{d} = 4$ is about the same as that for the case with $f_\mathrm{d} = 1$. As the physical origin of the decline of $\rho _\mathrm{d, KH}$ for $f_\mathrm{d} > 2$, we consider the difference of property of gas drag. After dust grains grow, the law of gas drag changes from Epstein's law to Stokes' law. Figure \ref{fig:12} shows the mass function at the onset of KHI at the maximum density for the case with $f_\mathrm{d} = 1$ and $f_\mathrm{d} = 4$, respectively. For the case with $f_\mathrm{d} = 1$, the mass function has the peak at $m / m_0 \sim 10^{12}$ and the typical size of dust grains is 0.9 cm. On the other hand, for the case with $f_\mathrm{d} = 4$, the mass function has the peak at $m / m_0 \sim 10^{14}$ and the typical radius of dust grains is 4 cm. In the case with $s \ge 3 l_\mathrm{g} / 2 = 2.2$ cm, the stopping time is given by Stokes' law. By comparison of these results, it is suggested that the decrease of dust density at KHI for $f_\mathrm{d} > 2$ in Figure \ref{fig:11} originates from the change of the law of gas drag due to dust growth.

\begin{figure}
  \begin{center}
    \FigureFile(74mm,74mm){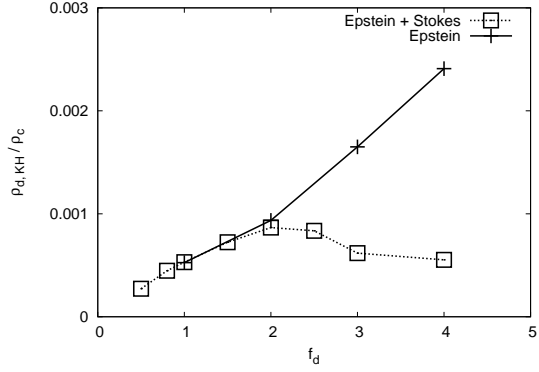}
  \end{center}
  \caption{The modified dust density in $r = 1$ AU at the onset of KHI for the case with dust growth in cases of using both Epstein's law and Stokes' law (dotted line and squares), and using only Epstein's law for all size (solid line and crosses). The abscissa and the ordinate show the same as ones of Figure \ref{fig:04}.}\label{fig:13}
\end{figure}

To confirm this possibility, for reference, we recalculate the evolution using Epstein's law for all size. Solid line in Figure \ref{fig:13} shows the dust density at the onset of KHI for this case. It is clearly seen that the dust density at the onset of KHI increases with increasing $f_\mathrm{d}$. By comparing two lines in Figure \ref {fig:13}, it is clear that the physical origin for the decline of $\rho _\mathrm{d, KH}$ for $f_\mathrm{d} > 2$ is the change of gas drag from Epstein's law to Stokes' law. Therefore, it is significant for us to take into account the dust size dependence of the stopping time as well as dust growth when we investigate the shear-driven turbulence in the protoplanetary disk.

\section{Discussion}

\subsection{Dependence on the heliocentric distance}

We estimate the dependence of the required abundance of dust for GI on the heliocentric distance $r$. First, we consider the case without dust growth. For simplicity, we consider the case when all dust grains have a single size. The scale height of dust density profile at the onset of KHI can be obtained from equation (\ref{eq:30}), and we define the scale height as $H_\mathrm{KHI}$. With substituting $\mathrm{Ri}_\mathrm{c}$ and $H_\mathrm{KHI}$ for $\mathrm{Ri} (z = z_\mathrm{c})$ and $H_\mathrm{d}$ in equation (\ref{eq:30}), respectively, $H_\mathrm{KHI}$ is given by
\begin{equation}
H_\mathrm{KHI} = \left( \frac{8 \mathrm{Ri}_\mathrm{c}}{27} \right) ^{\frac{1}{2}} \eta r = 1.0 \times 10^{-2} H_\mathrm{g} \left( \frac{r}{1 [\mathrm{AU}]} \right)^{\frac{1}{4}} \mathrm{.} \label{eq:43}
\end{equation}
Equation (\ref{eq:43}) shows that $H_\mathrm{KHI}$ is independent on $f_\mathrm{d}$. The dependence of $H_\mathrm{c}$ (defined in equation (\ref{eq:27})) on $r$ is given by $H_\mathrm{c} = 1.6 \times 10^{-5} f_\mathrm{d} \, \xi _\mathrm{ice} \, H_\mathrm{g} (r / 1 {~} [\mathrm{AU}]) ^{1/4}$, and we have
\begin{equation}
\frac{H_\mathrm{KHI}}{H_\mathrm{c}} = 6.6 \times 10^2 \left( \frac{f_\mathrm{d}}{1} \right) ^{-1} \left( \frac{\xi _\mathrm{ice}}{1} \right) ^{-1} \mathrm{.} \label{eq:44}
\end{equation}
The parameter of condensed water ice $\xi _\mathrm{ice}$ is given by 1 (at $r < 2.7$ AU) or 4.2 (at $r > 2.7$ AU) (\cite{key-H81}; \cite{key-H85}). Equation (\ref{eq:44}) shows the possibility that the required abundance of dust for GI is smaller outside the snow line than inside. Next, we consider the case with dust growth. At the onset of KHI, most of the dust grains with the typical size calculated in \S 4.3 (defined in equation (\ref{eq:37})) have settled near the midplane. Thus, the typical size of dust at the onset of KHI can be obtained as the size of dust grains which have settled from high altitudes to the midplane. As \citet{key-S69}, we have $dm = 4 \pi s^2 \rho _\mathrm{s} ds \sim - p_\mathrm{s} \pi (2 s)^2 \rho _\mathrm{d} (z) dz$. Assuming that $s = s_0$ at $z = + \infty$ and that $s = s_\mathrm{f} \gg s_0$ at $z = 0$, we derive $s_\mathrm{f}$ as
\begin{equation}
s_\mathrm{f} \sim \frac{p_\mathrm{s} \Sigma _\mathrm{d}}{2 \rho _\mathrm{s}} \sim \left( \frac{f_\mathrm{d}}{1} \right) \left( \frac{\xi _\mathrm{ice}}{1} \right) \left( \frac{r}{1 [\mathrm{AU}]} \right)^{-\frac{3}{2}} [\mathrm{cm}] \mathrm{.} \label{eq:45}
\end{equation}
The typical sizes of dust at the onset of KHI calculated in \S 4.3 were 0.9 cm and 4 cm in the case with $f_\mathrm{d} = 1$ and $f_\mathrm{d} = 4$, respectively, and consistent with equation (\ref{eq:45}). Equation (\ref{eq:45}) shows that the typical size of dust at the onset of KHI is smaller for larger $r$. On the other hand, the mean free path of gas molecules scales as $l_\mathrm{g} \propto {\rho _\mathrm{g}}^{-1} \propto H_\mathrm{g} {\Sigma _\mathrm{g}}^{-1} \propto r^{11/4}$, and $l_\mathrm{g}$ is larger for larger $r$. Thus, the gas drag force tends to be given by Epstein's law in the outer region of protoplanetary disk even for a large $f_\mathrm{d}$. The Solid line in Figure \ref{fig:13} shows that the dust density at the onset of KHI is larger for larger $f_\mathrm{d}$ in the Epstein regime. Therefore, the outer region of the protoplanetary disk might be more suitable for GI than the inner region.

Above discussion is different from that in \citet{key-T12} that showed that the inner part of the protoplanetary disk is more suitable for GI. \citet{key-T12} is based on the different condition from ours. \citet{key-T12} used the scale height of the dust layer which is determined by the balance between sedimentation and diffusion of dust grains, while we used the scale height of dust density profile at the onset of KHI.

\subsection{The linear analysis of the shear instability}

In this paper, we used Richardson number as an indicator of KHI in order to discuss the possibility of the shear-driven turbulence. However, although KHI occurs, there is a possibility that some dust grains continue to settle toward the midplane if the shear-induced turbulence is weak. An analysis with only Richardson number is insufficient in order to understand the effect of the shear-induced instability in the protoplanetary disk. \citet{key-S01} and \citet{key-M06} calculated the growth rate of the shear-induced instability by solving the linear perturbation equations. In these studies, the growth rate of the shear-induced instability depends on the assumed distribution of dust density which is not consistent with the formation process of dust layer. Thus, it would be better to calculate the growth rate of KHI as well as RTI with the consistent density distribution that is given in this paper. This will be addressed in the forthcoming paper.

\subsection{Possibilities of the streaming instability and the fractal growth of dust}

\citet{key-Y05} and \citet{key-J07} showed that the dynamics in the midplane is dominated by the streaming instability. \citet{key-B10} showed that dust grains with $\tau _\mathrm{s} \equiv \Omega _\mathrm{K} t_\mathrm{stop} > 0.01$ trigger the streaming instability before KHI. In our model, at $r = 1$ AU, $\tau _\mathrm{s} > 0.01$ corresponds to $s > 2$ cm. For our calculations, for the case with dust growth and $f_\mathrm{d} > 2$, the typical size of dust grains at the maximum density is larger than 2 cm. Therefore, in the dust layer governed by Stokes' law, the streaming instability would occur before KHI and GI.

In this paper, we assumed compact and spherical dust grains. However, since dust grains grow due to dust-dust collisions, large dust grains are aggregates of small dust grains. Both laboratory and numerical experiments show that aggregates are not at all compact and spherical, but have a fluffy structure (\cite{key-W98}; \cite{key-K99}; \cite{key-O09b}). With the same mass, the radius of the fractal aggregate is larger than the radius of the compact and spherical dust grain, and varies with the porosity of the aggregate. Therefore, it is essential to investigate the evolution of porosity during the growth of dust aggregate.

\section{Conclusion}

In this paper, we considered the sedimentation and the growth of dust grains, and we discussed the possibility of the shear-driven turbulence during dust sedimentation with and without dust growth. We assumed that the gas component is in hydrostatic equilibrium and that gas are not affected by motion of dust grains. Dust grains are assumed to be compact and spherical, and radial motion of dust grains are neglected. We used a single-fluid approximation for the azimuthal motion. For dust-dust collisions, it is assumed that dust grains collide and coalesce by sedimentation or the thermal Brownian motion.

Our study shows the following results: (1) The shear-driven turbulence is expected to occur before the onset of the gravitational instability in MMSN model at $r = 1$ AU. (2) In the case without dust growth, at the onset of KHI, dust density in the midplane is large for larger dust abundance. This tendency holds for the same initial size distribution of dust grains. (3) In the case with dust growth, dust density at the onset of KHI decreases with increasing dust abundance with $f_\mathrm{d} > 2$. This result is qualitatively different from the one in the case without dust growth. The reason is that gas drag changes from Epstein's law to Stokes' law for larger dust grains which grow up in advance. Thus, we stress that, for the study of shear-driven turbulence, the change of the law of gas drag from Epstein's law to Stokes' law as well as dust growth is required to be taken into account.

For the formation of planetesimals, it has been suggested that in order to occur GI before inward drift of dust, dust grains have to settle toward the midplane, and it is suggested to possible with large dust abundance. However, in this paper, it is suggested that the shear-driven turbulence is certain to happen even if dust abundance in the protoplanetary disk is larger than MMSN model. In future work, we should develop a more realistic model with calculating the effect of KHI and RTI directly.

\bigskip

\noindent We thank Fumio Takahara for fruitful discussion and continuous encouragement. We also acknowledge discussion with Sugawara in the early stage of this work.

\appendix
\section*{The typical radius of dust grains in the case with dust growth}

We explain the reason that the typical radius of dust grains $\bar{s} (z)$ is the linear function of $z$ at $z / H_\mathrm{g} \ll 1$ and at $t / t_\mathrm{sed} \ll 1$ owing to collisions due to sedimentation in the case with growth of dust grains. We now estimate the mean collision time in the same method used in \citet{key-N81}. The mean collision time is given by
\begin{equation}
t_\mathrm{coll} = \frac{1}{n_\mathrm{d} \sigma \Delta v} \mathrm{,} \label{eq:A01}
\end{equation}
where $n_\mathrm{d}$ is the number density of dust grains, $\sigma $ is the collisional cross section and $\Delta v$ is the relative velocity of the dust-dust collision. We assume that radii of dust grains are given by the typical radius and that masses of dust grains are given by the typical mass of dust grains. The typical mass is given by
\begin{equation}
\bar{m}(z) = \frac{4}{3} \pi \rho _\mathrm{s} [\bar{s}(z)]^3 \mathrm{,} \label{eq:A02}
\end{equation}
and we regard that $\sigma = \pi {\bar{s}}^2$. We treat $n_\mathrm{d}$ as $\rho _\mathrm{d} / \bar{m}$ or $(\Sigma _d / \Sigma _g) [\rho _\mathrm{g} / \bar{m}]$ at $z / H_\mathrm{g} \ll 1$ and at $t / t_\mathrm{sed} \ll 1$. For the collision due to sedimentation, we simply put $|s - s'| = \bar{s}$ and $\Delta v = \Delta v_\mathrm{s}$. The mean collision time for sedimentation is defined as $t_\mathrm{coll, \, s}$. At $z / H_\mathrm{g} \ll 1$ and at $t / t_\mathrm{sed} \ll 1$, $t_\mathrm{coll, \, s}$ is obtained by
\begin{equation}
t_\mathrm{coll, \, s} = \frac{2 \sqrt{2}}{3} \frac{\Sigma _\mathrm{g}}{\Sigma _\mathrm{d}} \frac{1}{\Omega _\mathrm{K}} \left( \frac{z}{H_\mathrm{g}} \right)^{-1} = 36 \left( \frac{z}{H_\mathrm{g}} \right)^{-1} [\mathrm{year}] \mathrm{.} \label{eq:A03}
\end{equation}
In equation (\ref{eq:A03}), it is seen that $t_\mathrm{coll, \, s}$ is independent of $\bar{s}$ and $t_\mathrm{coll, \, s} \propto z^{-1}$ at $z / H_\mathrm{g} \ll 1$ and at $t / t_\mathrm{sed} \ll 1$. For the collision due to the thermal motion, we simply put $m = m' = \bar{m}$ and $\Delta v = \Delta v_\mathrm{B}$. The mean collision time for the thermal motion, $t_\mathrm{coll, \, B}$, is obtained by
\begin{equation}
t_\mathrm{coll, \, B} = \frac{4}{3} \sqrt{\frac{2}{3}} \frac{\pi H_\mathrm{g}}{\Sigma _\mathrm{d}} \frac{{\rho _\mathrm{s}}^\frac{3}{2} {s_0}^{\frac{5}{2}}}{\sqrt{k_\mathrm{B} T}} \left( \frac{\bar{s}}{s_0} \right)^\frac{5}{2} \exp \left[ \left( \frac{z}{H_\mathrm{g}} \right)^2 \right] = 28 \left( \frac{\bar{s}}{s_0} \right)^\frac{5}{2} [\mathrm{year}] \mathrm{,} \label{eq:A04}
\end{equation}
at $z / H_\mathrm{g} \ll 1$ and at $t / t_\mathrm{sed} \ll 1$. In equation (\ref{eq:A04}), it is seen that $t_\mathrm{coll, \, B}$ is independent of $z$ and $t_\mathrm{coll, \, B} \propto {\bar{s}}^{\, 5 / 2}$ at $z / H_\mathrm{g} \ll 1$ and at $t / t_\mathrm{sed} \ll 1$.

From equations (\ref{eq:A03}) and (\ref{eq:A04}), it is found that $t_\mathrm{coll, \, B} < t_\mathrm{coll, \, s}$ while radii of dust grains are relatively small. Therefore, it is suggested that collisions due to the thermal motion is dominant as long as radii of dust grains are relatively small. After dust grains have grown, it is expected that $t_\mathrm{coll, \, B} > t_\mathrm{coll, \, s}$, and that collisions due to sedimentation are dominant. The growing speed of dust grains in sedimentation is expected to be proportional to $z$ because $t_\mathrm{coll, \, s} \propto z^{-1}$. Thus, it is supposed that the typical mass of dust grains at $z$, $\bar{m}(z)$, is given by $\bar{m}(z) = [a_1 (z / H_\mathrm{g}) + a_2] m_0$, where $a_1$ and $a_2$ are appropriate values. Then, the typical radius of dust grains at $z$ is given by $\bar{s}(z) = [a_1 (z / H_\mathrm{g}) + a_2]^{1 / 3} s_0$. If $a_1 z / a_2 H_\mathrm{g} \ll 1$, it is supposed that $\bar{s}(z)$ is approximated by $[a_3 (z / H_\mathrm{g}) + a_4] s_0$ with Taylor expansion. Symbols $a_3$ and $a_4$ are appropriate values.

\begin{figure}
  \begin{center}
    \FigureFile(74mm,74mm){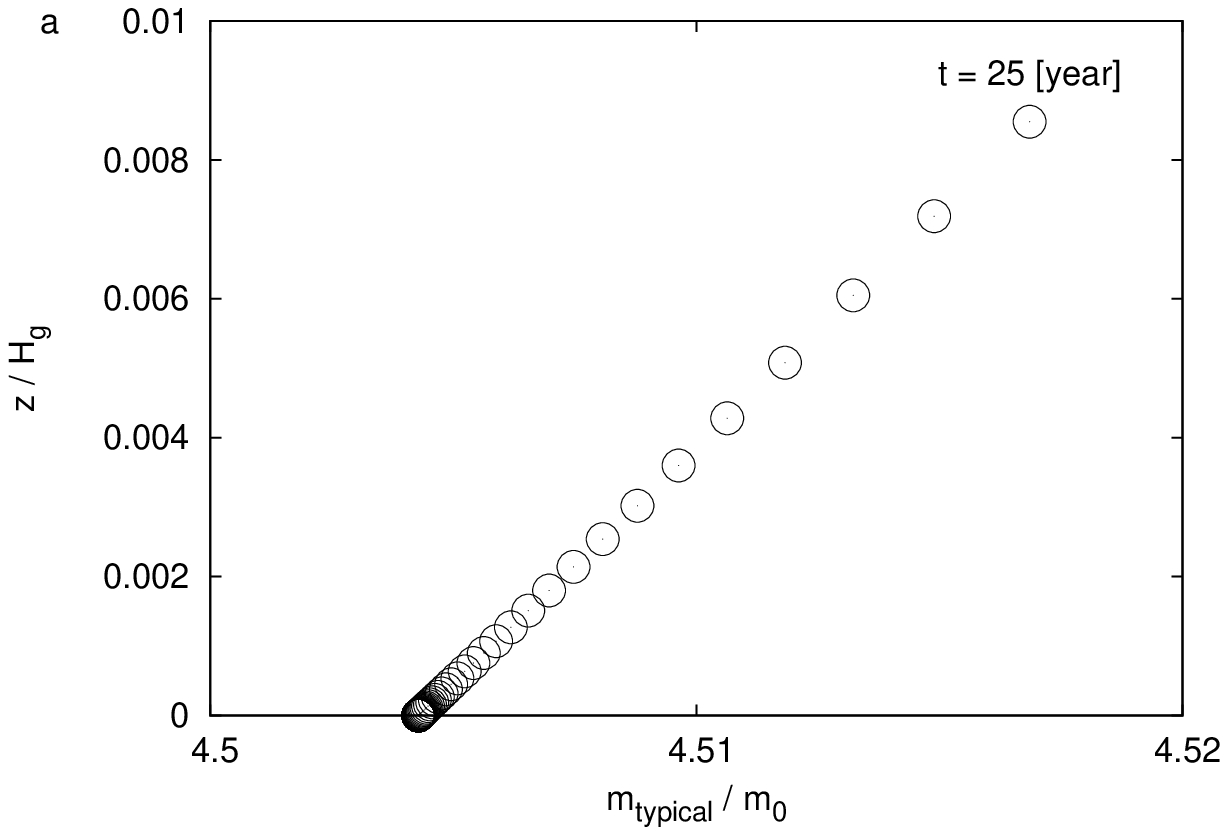}
    \FigureFile(74mm,74mm){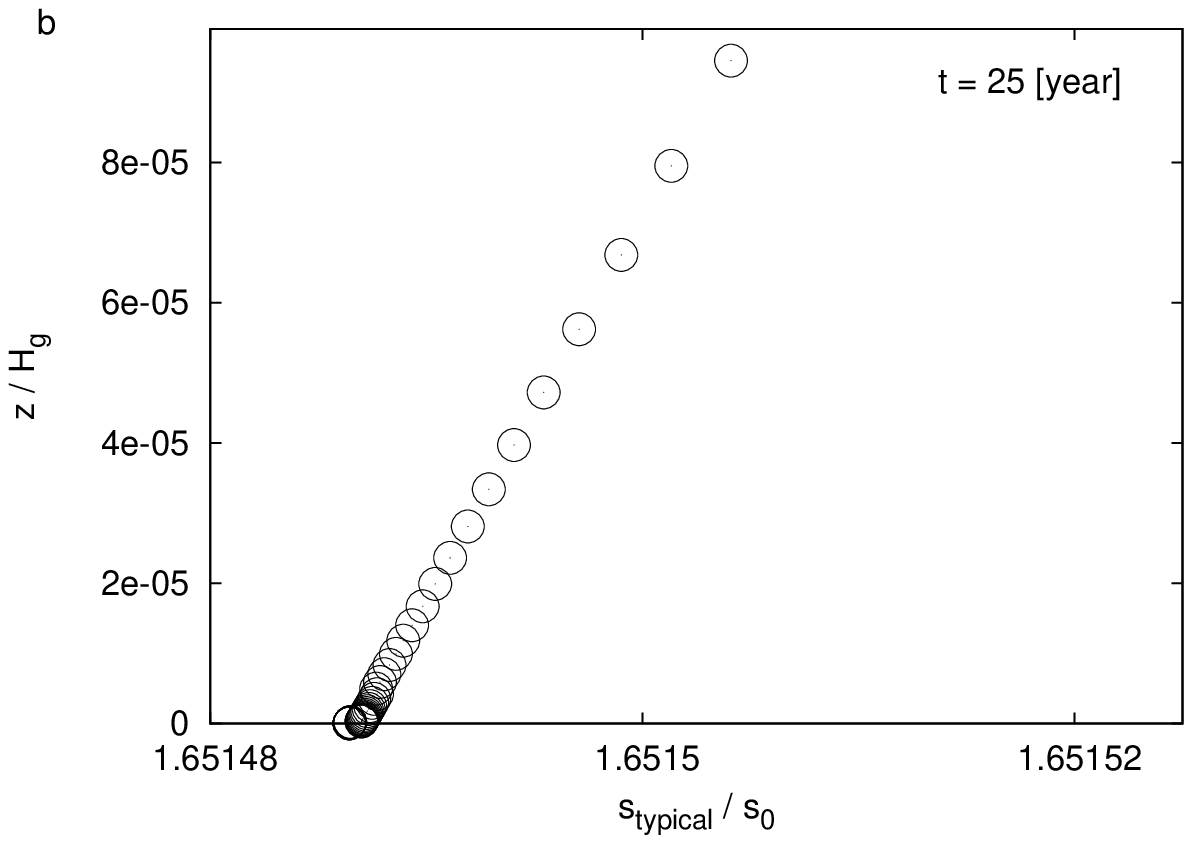}
  \end{center}
  \caption{(a) The distribution of the typical mass of dust grains in $r = 1$ AU at $t = 25$ year. The abscissa, where $m_\mathrm{typical}$ means $\bar{m}$, shows the typical mass $\bar{m}$ in unit of $m_0$. The ordinate shows $z$ coordinates in unit of $H_\mathrm{g}$. (b) The distribution of the typical radius of dust grains at $t = 25$ year. The abscissa, where $s_\mathrm{typical}$ means $\bar{s}$, shows the typical radius $\bar{s}$ in unit of $s_0$. The ordinate shows $z$ coordinates in unit of $H_\mathrm{g}$.}\label{fig:14}
\end{figure}

Figure \ref{fig:14} shows the distribution of the typical mass and radius of dust grains at $z / H_\mathrm{g} \ll 1$ and at $t = 25$ year. From Figure \ref{fig:14}, it is confirmed that $\bar{s}(z) = [a_3 (z / H_\mathrm{g}) + a_4] s_0$ at $z / H_\mathrm{g} \ll 1$ and at $t / t_\mathrm{sed} \ll 1$.

Above discussions assume that collisions due to sedimentation become more dominant than those due to the thermal motion immediately. However, it is not confirmed that collisions due to sedimentation is dominant to growth of dust grains before $t = 25$ year. We now should confirm that this assumption is appropriate for the case that we investigate.

At $z / H_\mathrm{g} \ll 1$ and $t \approx 0$, it is expected that collisions due to the thermal motion is dominant and $t_\mathrm{coll, \, B}$ is independent of $z$. In this case, it is supposed that the typical radius of dust grains is independent of $z$, i.e., $a_3$ in the formula for $\bar{s}(z)$ is temporally constant. However, in a certain time, it is expected that the dominant effect in the growth of dust grains changes from collisions due to the thermal motion to those due to sedimentation because of dust growth. Therefore, the time for this change can be determined by investigating the time development of $a_3$ for $\bar{s}(z)$.

\begin{figure}
  \begin{center}
    \FigureFile(74mm,74mm){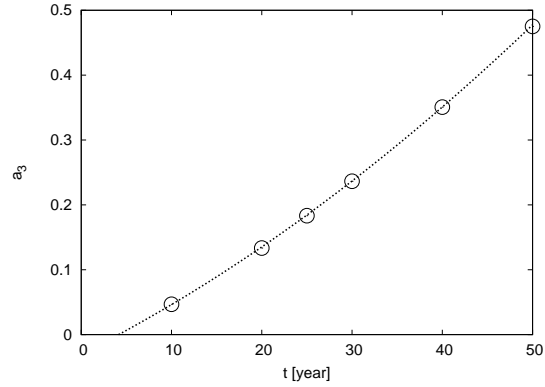}
  \end{center}
  \caption{The time development of $a_3$ (Circles) in $r = 1$ AU and $z / H_\mathrm{g} \ll 1$ at $t / t_\mathrm{sed} \ll 1$. The abscissa shows time and the ordinate shows $a_3$ that is derived from fitting $\bar{s}(z)$ into $[a_3 (z / H_\mathrm{g}) + a_4] s_0$. The approximated curve is also drawn (dotted line).}\label{fig:15}
\end{figure}

Figure \ref{fig:15} shows the time development of $a_3$. From Figure \ref{fig:15}, $a_3$ is approximated by $a_3 = 6.3 \times 10^{-5} (t / 1 {~} [\mathrm{year}])^2 + 7.0 \times 10^{-3} (t / 1 {~} [\mathrm{year}]) - 3.0 \times 10^{-2}$. This shows that $a_3 > 0$ at $t \gtrsim 4$ year, so it is considered that collisions due to sedimentation dominate in growth of dust grains at $t \gtrsim 4$ year. Therefore, we show that the assumption that collisions due to sedimentation become more dominant to growth of dust grains than those due to the thermal motion immediately is proper to the case that we investigate.

\begin{thebibliography}{}
\bibitem[Adachi et al.(1976)]{key-A76}
  Adachi,~I., Hayashi,~C., \& Nakazawa,~K. 1976, Prog.~Theor.~Phys., 56, 1756
\bibitem[Bai \& Stone(2010)]{key-B10}
  Bai~X., \& Stone~J.~M. 2010, \apj, 722, 1437
\bibitem[Chandrasekhar(1961)]{key-C61}
  Chandrasekhar,~S. 1961, Hydrodynamic \& Hydromagnetic Stability (New York: Dover)
\bibitem[Cuzzi et al.(1993)]{key-C93}
  Cuzzi,~J., Dobrovolskis,~A., \& Champney,~J. 1993, Icarus, 106, 102
\bibitem[Dobrovolskis(1999)]{key-D99}
  Dobrovolskis,~A.~R., Dacles-Mariani,~J.~S., \& Cuzzi,~J.~N. 1999, \jgr, 104, 30805
\bibitem[Garaud \& Lin(2004)]{key-G04}
  Garaud,~P., \& Lin,~D.~N.~C. 2004, \apj, 608, 1050
\bibitem[Goldreich \& Ward(1973)]{key-G73}
  Goldreich,~P., \& Ward,~W. 1973, \apj, 183, 1051
\bibitem[G\'{o}mez \& Ostriker(2005)]{key-G05}
  G\'{o}mez,~G.~C., \& Ostriker,~E.~C. 2005, \apj, 630, 1093
\bibitem[Hayashi(1981)]{key-H81}
  Hayashi,~C. 1981, Prog.~Theor.~Phys.~Suppl., 70, 35
\bibitem[Hayashi et al.(1985)]{key-H85}
  Hayashi,~C., Nakazawa,~K., \& Nakagawa,~Y. 1985, in Protostars \& Planets II, ed.\ D.~C.~Black \& M.~S.~Mathews (Tuscon: Univ. Arizona Press), 1100
\bibitem[Johansen et al.(2006)]{key-J06}
  Johansen,~A., Henning,~Th., \& Klahr,~H. 2006, \apj, 643, 1219
\bibitem[Johansen \& Youdin(2007)]{key-J07}
  Johansen,~A., \& Youdin,~A. 2007, \apj, 662, 627
\bibitem[Kempf et al.(1999)]{key-K99}
  Kempf,~S., Pfalzner,~S., \& Henning,~T.~K. 1999, Icarus, 141, 388
\bibitem[Michikoshi \& Inutsuka(2006)]{key-M06}
  Michikoshi,~S., \& Inutsuka,~S. 2006, \apj, 641, 1131
\bibitem[Nakagawa et al.(1981)]{key-N81}
  Nakagawa,~Y., Nakazawa,~K., \& Hayashi~C. 1981, Icarus, 45, 517
\bibitem[Nakagawa et al.(1986)]{key-N86}
  Nakagawa,~Y., Sekiya,~M., \& Hayashi~C. 1986, Icarus, 67, 375
\bibitem[Okuzumi(2009)]{key-O09a}
  Okuzumi,~S. 2009, \apj, 698, 1122
\bibitem[Okuzumi et al.(2009)]{key-O09b}
  Okuzumi,~S., Tanaka,~H., \& Sakagami,~M-a. 2009, \apj, 707, 1247
\bibitem[Roe(1986)]{key-R86}
  Roe,~P.~L. 1986, Ann.~Rev.~Fluid Mech., 18, 337
\bibitem[Safronov(1969)]{key-S69}
  Safronov,~V.~S. 1969, Evolution of the Protoplanetary Cloud \& the Formation of the Earth \& Planets (Moscow: Nauka Press)
\bibitem[Sekiya(1983)]{key-S83}
  Sekiya,~M. 1983, Prog.~Theor.~Phys., 69, 1116
\bibitem[Sekiya(1998)]{key-S98}
  Sekiya,~M. 1998, Icarus, 133, 298
\bibitem[Sekiya \& Ishitsu(2001)]{key-S01}
  Sekiya,~M., \& Ishitsu,~N. 2001, Earth Planets Space, 53, 761
\bibitem[Takeuchi et al.(2012)]{key-T12}
  Takeuchi,~T., Muto,~T., Okuzumi,~S., Ishitsu,~N., \& Ida,~S. 2012, \apj, 744, 101
\bibitem[Trubnikov(1971)]{key-T71}
  Trubnikov,~B.~A. 1971, Sov.~Phys.~Dokl., 16, 124
\bibitem[Weidenschilling(1980)]{key-W80}
  Weidenschilling,~S.~J. 1980, Icarus, 44, 172
\bibitem[Wetherill \& Stewart(1989)]{key-W89}
  Wetherill,~G.~W., \& Stewart,~G.~R. 1989, Icarus, 77, 330
\bibitem[Wurm \& Blum(1998)]{key-W98}
  Wurm,~G., \& Blum,~J. 1998, Icarus, 132, 125
\bibitem[Youdin \& Goodman(2005)]{key-Y05}
  Youdin,~A.~N., \& Goodman,~J. 2005, \apj, 620, 459
\bibitem[Youdin \& Shu(2002)]{key-Y02}
  Youdin,~A.~N., \& Shu,~F.~H. 2002, \apj, 580, 494
\bibitem[Zsom et al.(2010)]{key-Z10}
  Zsom,~A., Ormel,~C.~W., G\"{u}ttler,~C., Blum,~J., \& Dullemond,~C.~P. 2010, \aap, 513, A57
\end{thebibliography}
\end{document}